\documentclass[twocolumn,showpacs,preprintnumbers,amsmath,amssymb,superscriptaddress,prb,english]{revtex4-1}
\usepackage{epsfig,amsopn}
\usepackage{amssymb,epsfig}
\usepackage{epsfig}
\usepackage{graphicx}
\graphicspath{{figure/}}
\usepackage{epstopdf}
\usepackage{color}
\usepackage[caption=false]{subfig}
\usepackage{amsmath,amssymb}
\usepackage{amsthm}
\usepackage{enumerate}
\usepackage[T1]{fontenc}
\usepackage[utf8]{inputenc}
\usepackage{mathtools}   % loads »amsmath«
\usepackage{empheq}
\usepackage{physics}
\usepackage{cancel}

\newcommand{\comment}[1]{}
\newcommand\beq{\begin{equation}}
\newcommand\eeq{\end{equation}}

\begin{document}
\title{Fermi Arc Reconstruction at the Interface of Twisted Weyl Semimetals}

 \author{Faruk Abdulla}
 \affiliation{Harish-Chandra Research Institute, HBNI, Chhatnag Road, Jhunsi, Allahabad 211 019, India}
\author{ Sumathi Rao}
 \affiliation{Harish-Chandra Research Institute, HBNI, Chhatnag Road, Jhunsi, Allahabad 211 019, India}
 \author{Ganpathy Murthy}
 \affiliation{Department of Physics and Astronomy, University of Kentucky, Lexington, KY 40506-0055}

\begin{abstract}

Three-dimensional Weyl semimetals have pairs of topologically
protected Weyl nodes, whose projections onto the surface Brillouin
zone are the end points of zero energy surface states called Fermi
arcs. At the endpoints of the Fermi arcs,  surface states extend into
and are hybridized with the bulk. Here, we consider a two-dimensional
junction of two identical Weyl semimetals whose surfaces are twisted
with respect to each other and tunnel-coupled. Confining ourselves to
commensurate angles (such that a larger unit cell preserves a reduced
translation symmetry at the interface) enables us to analyze arbitrary
strengths of the tunnel-coupling.  We study the evolution of the Fermi
arcs at the interface, in detail, as a function of the twisting angle
and the strength of the tunnel-coupling. We show unambiguously that in
certain parameter regimes, all surface states decay exponentially into
the bulk, and the Fermi arcs become Fermi loops without endpoints. We
study the evolution of the `Fermi surfaces' of these surface states as
the tunnel-coupling strengths vary.  We show that changes in the
connectivity of the Fermi arcs/loops  have interesting signatures in the
optical conductivity in the presence of a magnetic field perpendicular
to the surface.

\end{abstract}

\maketitle

%-----------------------------------------------------------------------------------------------------------------------------------------------------------------%-----------------------------------------------------------------------------------------------------------------------------------------------------------------

\textcolor{magenta}{\section{Introduction}}
 \label{sec:intro}

Weyl
semimetals \cite{Murakami2007,Wan2011,Yang2011,Burkov2011,Xu2011,Potter2014,Xu2015,Lv2015,Lu2015,Moll2016}
are often described as three dimensional analogues of
graphene \cite{graphene}, with band-touchings or nodes at isolated
points in the Brillouin zone. These nodes are chiral, and can be
obtained by separating the Dirac nodes of a three dimensional
semimetal by either time-reversal \cite{Murakami2007} or inversion
symmetry \cite{Wan2011,Yang2011,Xu2011,Burkov2011} breaking.  The low
energy excitations about these nodes are Weyl fermions with
anisotropic velocities that depend on the material parameters. Weyl
semimetals (WSMs) exhibit several novel features such as negative
longitudinal magneto-resistance \cite{Wan2011,Son2013}, anomalous Hall
effect \cite{Xu2011,Yang2011}, chirality dependent Hall
effect \cite{Yang2015}, planar Hall effect \cite{Nandy2017}, etc.

Several unconventional features have also been uncovered by studying
transport across junctions of these Weyl semimetals with other
topological and non-topological materials.  For instance, junctions of
Weyl semi-metals with superconductors have also led to new phenomena
such as chirality dependent Andreev
reflection \cite{Ueda2014,Bovenzi2017} and chirality dependent
Josephson effects. \cite{Khanna2016,Khanna2017}  Tunneling conductances
across WSM-barrier-WSM junctions have also been studied with interesting
experimentally testable
consequences. \cite{Mukherjee2017,Mukherjee2019,Sinha2019}

The band topology of the WSM is encoded in the monopole charge or the
Chern number of the Berry curvature carried by the Weyl node. Hence
surfaces in the bulk Brillouin zone (BZ) which enclose only one of the
nodes carry Chern number.  This leads to surface states called Fermi
arc (FA) states in the surface BZ joining the projections of the Weyl
nodes on to the surface BZ. Since the end-points of the FAs are the
projections of the Weyl nodes, FAs on one surface connect to the FAs
on the opposite surface through the bulk nodes. In the presence of a
small magnetic field, this gives rise to intersurface cyclotron orbits
\cite{Potter2014} which depend on the thickness of the sample. These
exotic FA states are the hallmark of Weyl semimetals and it was their
initial experimental identification \cite{Xu2015} using angle-resolved
photoemission spectroscopy that  led to the current explosion in
theoretical interest \cite{Nonlocal2015,Deb2017,McCormick2018} in
understanding their properties.  More recently, Shubnikov-de Haas
oscillations \cite{Moll2016} and the quantum Hall effect based on
intersurface cyclotron orbits \cite{3DQHE_WSM} have been seen in
$Cd_3As_2$.

Our aim in this work is to study the physics that emerges when two
slabs of WSM are twisted with respect to each other and
tunnel-coupled. From the analogy to graphene bilayers
\cite{Novoselov2016,Duong2017,Mele2010,Mele2011} which show
interesting effects, including the emergence of highly correlated
states when the two layers have a small ``magic angle'' twist with
respect to each other, \cite{Bistritzer2011,Sanjose2012} we might
expect new physics, both in the bulk of the WSMs and in the interface
FA states.

The WSM-WSM junction with no twist was initially studied in
Refs. \onlinecite{Dwivedi2018,Ishida2018} where FA reconstructions
were found when the junction was between WSMs with different FA
connectivities. Coupling of WSMs with small {\it incommensurate}
twists has also been studied earlier by Murthy, Fertig, and Shimshoni
\cite{Murthy2020} (henceforth MFS), in a perturbative regime of
tunnel-coupling.  MFS showed that, due to the effective Moire
Brillouin zone that can be defined in terms of the mismatched lattice
wave-vectors, reconstructions of the FAs take place. They also
conjectured that at certain ``arcless angles'', at sufficiently strong
tunnel-coupling, the reconstructed surface states would consist of
Fermi loops totally disconnected from the projections of the Weyl
points on the surface BZ.

%They also explicitly studied the lattice model where the angle
%between the two twisted WSMs was fixed to be $\theta=\pi/2$ and
%$\theta=\pi/4$ for small tunnel couplings.

 In this paper, we extend MFS's work to arbitrary commensurate angles
 with a reduced lattice translation symmetry, and thus a larger
 superlattice unit cell at the interface. The presence of true lattice
 translation symmetry (absent in the work of MFS) allows us to analyze
 arbitrary strengths of the tunnel couplings between the two slabs.
 We perform a detailed study of the evolution of the FAs as a function
 of the coupling strength of the tunnel coupling for two
 sequences of commensurate twist angles parametrized by a
 positive integer $n$, $\theta_n=\tan^{-1}(1/n)$ and $\theta_n=\pi/2+\tan^{-1}(1/n)$.  We
 unambiguously show that there exist parameter regimes where all the 
 surface states are disconnected from the bulk, like the surface states of
 a topological insulator. We take a detailed look at the liftoff or
 detachment transition, where the Fermi arc detaches itself from the
 Weyl node projection and forms a surface state with a closed Fermi
 surface. We analyze the different `geometries' into which the closed
 Fermi surfaces evolve. Finally, we uncover a duality between strong
 and weak interface tunnel couplings.

The plan of this paper is as follows. In Section II, we define our
model Hamiltonian and the parameters that enter it, which include the
commensurate twist angle and the interface hopping matrix. In Section
III, we study the evolution of the interface FA states as a function
of the twist parameter and the strength of the tunnel-coupling for two
simple commensurate angles with the smallest superlattices at the
interface, which display nearly all the phenomena of interest. In
Section IV we present the simplest model of the liftoff or detachment
transition, where the FA detaches itself from the Weyl point
projection on the the surface BZ. We end in Section V with
conclusions, caveats, a discussion of potential experimental
signatures of the phenomena we uncover, and some promising future
directions. Many important details of the calculations for larger
interface superlattices, the symmetries of the model, the stability of
our conclusions to longer range tunnel-couplings, etc are relegated to
a series of appendices.

\begin{figure}
\includegraphics[width=0.9\linewidth, height=0.9\linewidth]{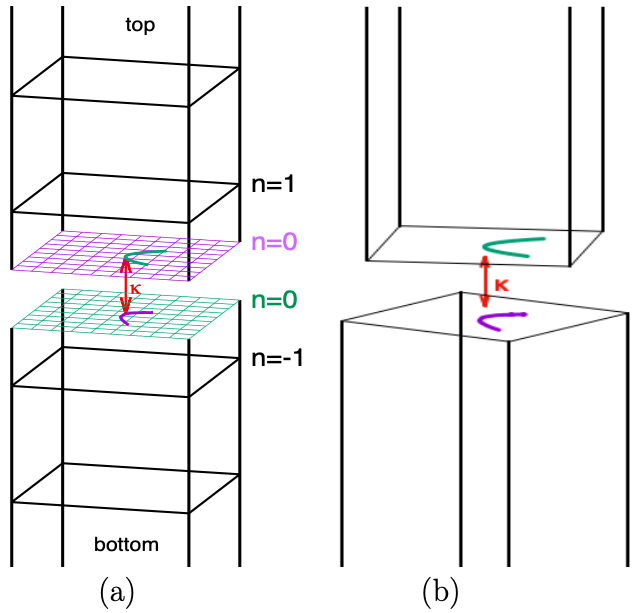}
\caption{a) Two identical semi infinite WSM slabs are tunnel coupled. The coupling is parametrised by $\kappa$. 
The interface (at $z=0$) consists of two $n=0$ layers, one from the top slab and the other from the bottom slab. (b) 
The two  slabs are twisted by $\theta$ clockwise and anticlockwise by the  same angle and then coupled.
For a commensurate twist angle $\theta$, there will be a superlattice(SL) at the interface (an example is shown in Fig. \ref{fig-SL}). }
\label{fig-slabs}
\end{figure}

%--------------------------------------------------------------------------------------------------------------------------------------------------------------
%--------------------------------------------------------------------------------------------------------------------------------------------------------------
\textcolor{magenta}{\section{Twisted Weyl Semimetals  and the Interface  Hamiltonian}}
\label{sec:TWSM}
\subsection{ Time-reversal symmetry broken model of a WSM and its surface states}
\label{subsec:STWSM}

To set the notation for the interface states of two twisted WSMs, we
briefly review the derivation of the surface states for a
semi-infinite WSM\cite{Murthy2020}.  We begin with a general two band
lattice model of a time reversal symmetry broken WSM on a cubic
lattice. The Hamiltonian  in momentum space is 
\begin{align}
H_0 = & \sum_{\bf k} c^{\dagger}_{\bf k} \{ 2t' \sin k_z \sigma_z + 2\sin k_y \sigma_y  \nonumber \\ &
  + 2 \sigma_x (2 + \cos k_0 - \cos k_x - \cos k_y - \cos k_z) \} c_{\bf k}~. \label{eq:ham}
\end{align}
The spectrum has Weyl nodes at $(\pm k_0, 0, 0)$ with chirality $\pm
1$ respectively.  Here $\sigma_{\mu}$, $\mu= (x, y, z)$ are the spin
Pauli matrices and $c_{\bf k} = (c_{{\bf k} \uparrow}, c_{{\bf k}
  \downarrow})^T$, are two component fermions. The hopping amplitude
within the $x$-$y$ plane has been set to unity and $t'$ represents the
hopping amplitude along the $z$ direction. We choose our units so that
the lattice constant $a$ can be set to unity throughout the paper.  In
this geometry, we expect Fermi arc states to be present on surfaces
which are not normal to the $x$-axis, that is, those with surface
Brillouin zones defined by $(k_x, k_y)$ or $(k_x, k_z)$.

To find the surface states, following Ref. \onlinecite{Murthy2020}, we
assume that the slab is semi-infinite in the $z$-direction, with
lattice sites labeled by $n=0,1,2 \dots$ and Fourier transform
$H_0({\bf k})$ to real space (in the $z$ direction) to get
\begin{align}
H_0(k_x, k_y) & =  \sum_n \{  2 c^{\dagger}_n (f_x\sigma_x + f_y\sigma_y) c_n - c^{\dagger}_{n}(\sigma_x + i t'\sigma_z)    \nonumber \\
& \hspace{0.6cm} \times c_{n+1}  - c^{\dagger}_{n+1} (\sigma_x - i t'\sigma_z) c_n  \}
\end{align}
where $ f_x = (2+\cos{k_0} - \cos{k_x} - \cos{k_y})$ and $f_y= \sin{k_y}$. We have suppressed the $k_x, k_y$ dependence  of all the fermion operators for notational simplicity. Following a rotation of the $\sigma$ matrices by $\pi/2$ around the $x$-axis, we get 
the transformed Hamiltonian to be
\begin{align}\label{eq-wsm0}
\tilde{H}_0(k_x, k_y) & =   \sum_{n=0}^{\infty} \{  2 c^{\dagger}_n (f_x\sigma_x + f_y\sigma_z) c_n - c^{\dagger}_{n}(\sigma_x - i t'\sigma_y)    \nonumber \\
& \hspace{0.6cm} \times c_{n+1} - c^{\dagger}_{n+1} (\sigma_x + i t'\sigma_y) c_n  \},  
\end{align}
which turns out to be real and hence, more convenient for further analysis. As shown in detail in Ref. \onlinecite{Murthy2020},
requiring the decaying solutions into the bulk to be normalizable, and assuming that $0<t' =\sin\phi$ and $k_0 <\phi$ 
gives  the dispersion of the surface states as $E(k_x, k_y) = 2f_y = 2\sin{k_y}$ and the eigenstates are spin-polarized along the $\sigma_z$ direction.

\subsection{Interface between two identical twisted WSMs}
\label{interface}

\begin{figure*}
\subfloat[ ]{\includegraphics[width=0.5\linewidth, height=0.35\linewidth]{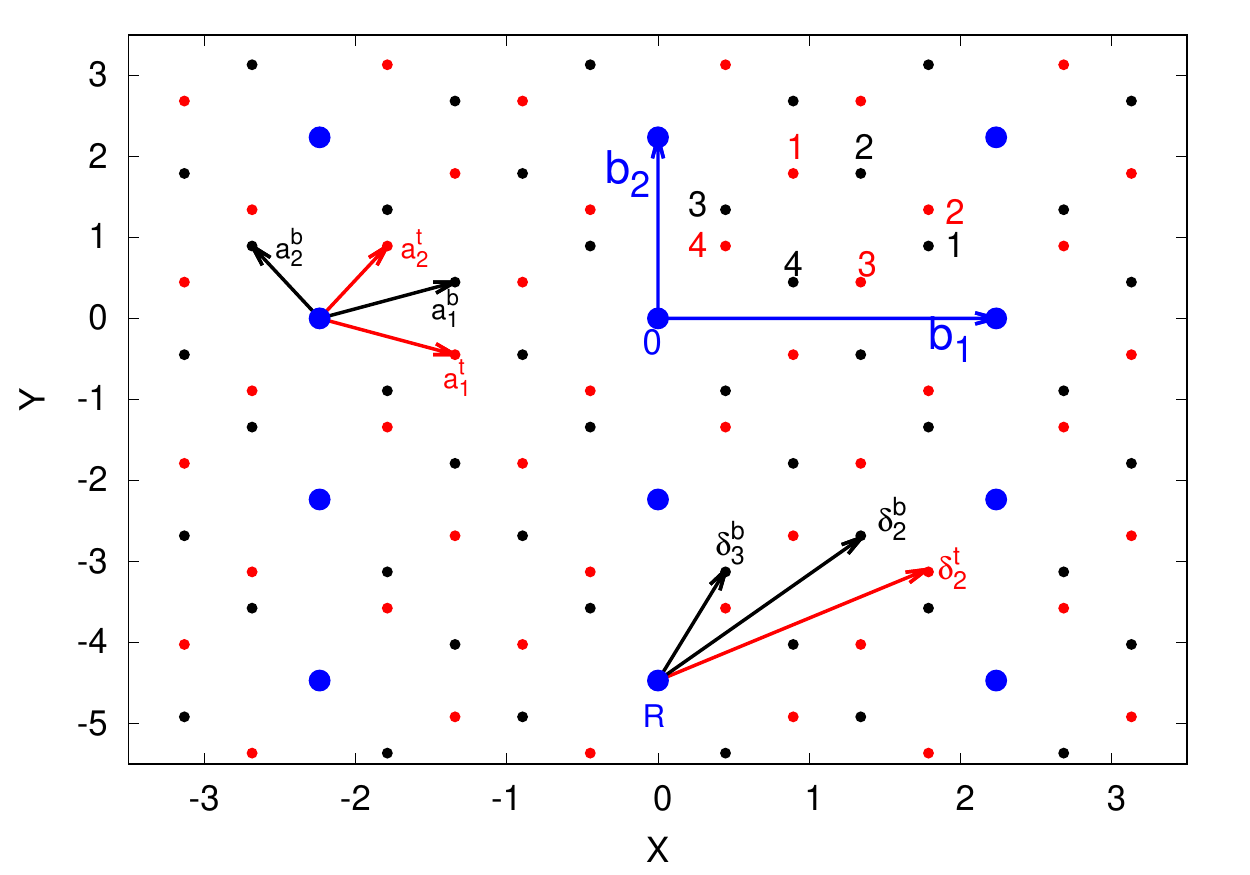}
\label{fig-SL}}
\subfloat[ ]{\includegraphics[width=0.45\linewidth, height=0.35\linewidth]{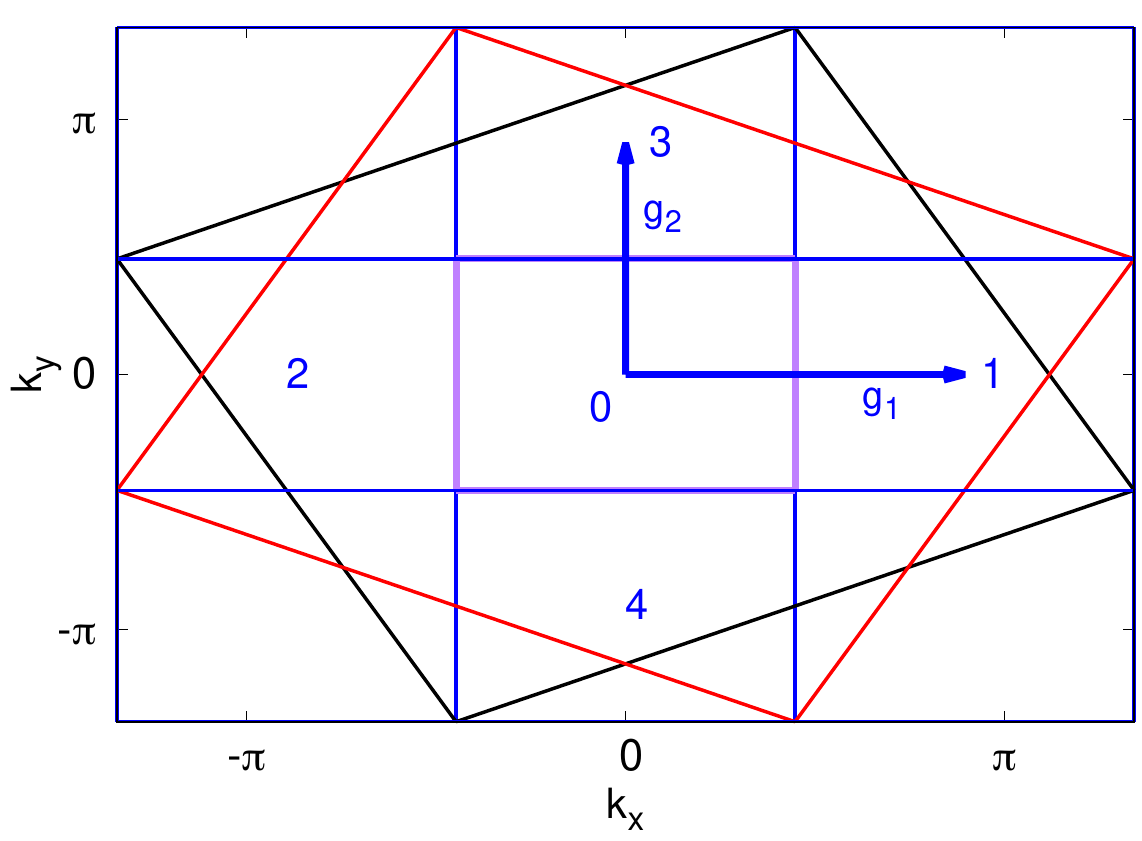}
\label{fig-SLbz}}
\captionsetup{position=bottom}
\caption{a) The interface layers for $n_0=2$. Here the  top and bottom  slabs are  rotated  by $\theta_2= \pm \tan^{-1}(1/2)$  so that the site $(2,1)$ of the $n=0$ layer of the top slab  coincides with the $(2, -1)$  site of the $n=0$  layer of the bottom slab. The lattice sites of the top and bottom slab layers
are in red and black colour respectively. The superlattice (SL) sites are in blue and ${\bf b}_1$, ${\bf b}_2$ are its primitive vectors.The SL unit cell contains $N_{sc}/2 = 2(n_0^2 +1)/2=5$  lattice sites of each slab and they are labelled by $\alpha=\{0, 1, 2, 3, 4\}$. (b) Here, the Brillouin zones of the rotated slabs and of the SL lattice are shown. In addition to the first SL BZ (labelled $0$), the second BZs of the SL are also shown with labels $1, 2, 3$ and $4$  and   ${\bf g}_1, {\bf g}_2$ are the reciprocal lattice vectors of the SL. For this particular rotation, the SL BZ  is $N_{sc}/2=5$  times smaller than the BZ of original  lattice.  }
\label{fig-SL-BZ}
\end{figure*}

In this subsection, we will see how the Fermi arcs of two identical
WSM slabs, twisted with respect to each other, get modified and
reconstructed when we switch on a tunnel-coupling between them.  We
consider the interface of two identical WSMs as shown in
Fig. \ref{fig-slabs}.  Both the WSMs are semi-infinite and the layers
of the top slab are labeled by $n=0, 1, 2, ..$ and the layers of the
bottom slab are labeled by $n=0, -1, -2, ..$. The interface consists
of the zeroth layer of both the slabs, which are tunnel-coupled after
twisting the top and bottom slabs around the z-axis by an angle
$\pm\theta_{n_0}$ in the clockwise direction, analogous to the
rotation of individual layers in bilayer graphene. \cite{Mele2010} The
subscript $n_0$ indicates that we rotate the top slab clockwise and
the bottom slab anticlockwise until the lattice site $(n_0,1)$ of the
top layer coincides with the lattice site $(n_0,-1)$ of the bottom
layer, where $n_0$ is a non negative integer. The twist angle in this
case is clearly $ \theta_{n_0}=\tan^{-1}(1/n_0)$. This results in a
periodic superlattice (SL) structure at the interface. We will
consider only such commensurate twists in this paper.

The SL unit cell contains an equal number of lattice sites from the
top layer and the bottom layer. For a given twist angle
$\theta_{n_0}$, there are a total $N_{sc} = 2\times(n_0^2 + 1)$
lattice sites per SL unit cell when $n_0$ is even (see
Fig. \ref{fig-SL} for an example when $n_0=2$). When $n_0$ is odd,
then\ total number of lattice sites per SL unit cell are $N_{sc} =
(n_0^2 + 1)$. For instance , when $n_0=0, 1, 2, 3, 4$ and $5$, the
explicit number of total lattice sites per SL unit cell are $N_{sc} =
2, 2, 10, 10, 34$ and $26$ respectively. The relative size (area) of
the SL unit cell at the interface (with respect to the original 2D
unit cell) is determined by $N_{sc}$ and it is actually $N_{sc}/2$
times larger.

In the presence of tunnel-coupling linking the top and bottom slabs,
the full Hamiltonian consists of
\begin{equation}
  H = H^t + H^b + \kappa H_V .
\end{equation}
where the $t$ and $b$ refers to the top and bottom slabs and both of
them are the same as $\tilde{H}_0(k_x, k_y)$ (Eq. \ref{eq-wsm0})
except that the slabs are now rotated with respect to each other and
$H_V$ is the coupling Hamiltonian (see
Fig. \ref{fig-slabs}). The overall strength of the
  coupling is parametrised by $\kappa$. The content of the
  spin-space matrix in $H_V$ will be described shortly. Since we started
with a cubic lattice, the planar lattice is a square and after
rotation, the primitive lattice vectors of the top layer are given by
${\bf a}^t_1=(\cos{\theta_{n_0}}, -\sin{\theta_{n_0}}), {\bf
  a}^t_2=(\sin{\theta_{n_0}}, \cos{\theta_{n_0}})$ and that of the
bottom layer are ${\bf a}^b_1=(\cos{\theta_{n_0}},
\sin{\theta_{n_0}}), {\bf a}^b_2=(-\sin{\theta_{n_0}},
\cos{\theta_{n_0}})$ as shown in Fig. \ref{fig-SL} (for $n_0=2$). The
Hamiltonian is now explicitly given by
\begin{align}\label{eq-slabHamiltonian}
H^\gamma = & \sum_{\bf k} \sum_{n} [ 2 c^{\dagger}_n({\bf k})  {\bf M }^\gamma c_n({\bf k}) - c^{\dagger}_{n+1}({\bf k}){\bf T} c_n({\bf k}) \nonumber \\ 
 &   \hspace{0.4cm} - c^{\dagger}_{n}({\bf k}) {\bf T}^{\dagger} c_{n+1}({\bf k}) ]
\end{align}
where ${\bf k}$ here refers to the transverse momentum vector $(k_x,k_y)$ and  $\gamma=(t, b)$ refer to the top and bottom layers,
and the sum over $n$ goes  from $n=0,1,2,\dots$ for the top layer and from $n=0,-1,-2,\dots$ for the bottom layer. The onsite
and hopping matrices are given by 
\begin{align} \label{hoppingmatrices}
{\bf M}^{\gamma}  &= (f_1^{\gamma} \sigma_x + f_3^{\gamma}\sigma_z ),  \nonumber  \\
  {\bf T} & = (\sigma_x + it' \sigma_y);  
 \end{align} 
 with 
 $f_1^{\gamma}({\bf k}) = 2 + \cos k_0 - \cos({\bf k.a}_1^{\gamma}) - \cos({\bf k.a}_2^{\gamma})$ and   $f_3^{\gamma} = \sin({\bf k.a}_2^{\gamma})$

 As mentioned earlier, the interface consists of  the zeroth layers of the WSM slabs, tunnel coupled by the hopping Hamiltonian $H_V$, given by 
\begin{align} \label{eq-coupling}
H_V =  \sum_{{\bf r}_t} \sum_{{\bf r}_b} c^{\dagger}_{t0s}({\bf r}_t) V_{ss'}(|{\bf r_t - r_b}|) c_{b0s'}({\bf r}_b) + h.c ,
\end{align}
where the lattice sites ${\bf r}_t$ and ${\bf r}_b$ live on the $n=0$
layer of the top and bottom slabs respectively.  Note that we have now
introduced the labels $t$ and $b$ to distinguish the fermions that
live on the top and bottom layers.  We have also introduced the spin
indices $s$ and $s'$ since the tunnel coupling is a matrix in spin
space.  We assume short-range hopping so that $V_{ss'}$ is nonzero
only for sites on the two surfaces with the same 2D coordinates.
\begin{equation}
  \label{eq-V-short} V_{ss'}(|{\bf
    r_t - r_b}|) = (V_0)_{ss'} \delta_{{\bf r}_t, {\bf r}_b} .
\end{equation}
These are the larger blue dots in Fig. \ref{fig-SL}.  In Appendix
\ref{App:LongRangeHopping}, we show that the perturbative inclusion of
longer-range hoppings does not qualitatively change the results that
we obtain in our model.

Note that the most general tunneling matrix between two sites  can be written as
a matrix in spin-space,
\beq
V = \sum_{i=0}^3 V_i \sigma_i,
\label{eq:tunneling}
\eeq
where $V_i$ are complex numbers, $\sigma_0$ is the identity
matrix and $\sigma_i$ are the three Pauli matrices for $i=x,y,z$.
This gives us a eight-parameter space of tunneling matrix elements,
which is difficult to explore systematically. 

Fortunately, there is a natural way to restrict the space of
tunnel-couplings. When all tunneling matrix elements between the slabs
are set to zero, our model (setting $\kappa=0$) enjoys a large number of symmetries, which
include unitary, anti-unitary, particle-particle, and particle-hole
type symmetries. For example, it is clear from Fig. \ref{fig-SL} that,
for $n_0=2$ the SL is symmetric about the positive diagonal with the
replacement of sites $i=1,..4$ of the upper layer with the
corresponding sites of the lower layer. It is also symmetric along the
negative diagonal, with the replacement of the sites $1
\leftrightarrow 3$ and $2 \leftrightarrow 4$ between the upper and
lower layers.  These symmetries are detailed in Appendix \ref{App:Symmetry}. In
general, not all the symmetries of the uncoupled model can be
satisfied by the tunneling term.  {We will assume that
  the tunneling conserves the symmetries of rotation by $\pi$ around
  both diagonals of the SL unit cell of the
  surface Brillouin zone for twist angles of the form $\theta_n=\pi/2 +
  \tan^{-1}(1/n)$ .  For   $\theta_n = \tan^{-1}(1/n)$, on the other hand, we will
  assume symmetry of rotation by $\pi$ around  the x and y axes. In both cases, this }leads to a
restriction that the tunneling matrix be real, and a combination of
$\sigma_x$ and $\sigma_y$ only. It is thus of the form $V_x\sigma_x+i
V_y\sigma_y$ where $V_x,V_y$ are real. We have verified in Appendix \ref{App:SymmetryPerturbation}
that perturbations violating these symmetries do not qualitatively
change our conclusions.

To physically understand why $\sigma_x$ and/or $\sigma_y$ terms must
be present in order to get a reconstructed Fermi arc at the interface,
recall that the surface states of the top/bottom slabs are spin
up/down polarized (see subsection \ref{subsec:STWSM}). So keeping
solely $\sigma_0$ and/or $\sigma_z$ terms cannot lead to any coupling
between the unreconstructed Fermi arcs of the top and bottom slabs.

\textcolor{magenta}{\section{Interface states and their evolution}}
\label{sec:result}

In this section, we will study the coupled FA states at the interface
of the two WSMs. These are states that are localised on the interface
and decay exponentially into the bulk. The individual FA eigenstates
are \cite{Murthy2020} (details reviewed in Appendix \ref{App:ComputationFermiArc})
\begin{align}
\ket{\Psi^{t}({\bf k})} &= \sum_{n=0}^{\infty} \psi^t_n({\bf k}) c^{\dagger}_n({\bf k}) \ket{0} \\
\ket{\Psi^{b}({\bf k})} &= \sum_{n=0}^{-\infty} \psi^b_n({\bf k}) c^{\dagger}_n({\bf k})\ket{0}.
\end{align}
where $\psi^t_n({\bf k}) \in \mathbb{C}^2 $ ($n \geq 0$) and
$\psi^b_n({\bf k}) \in \mathbb{C}^2 $ ($n \leq 0$) are wavefunctions,
living on the $n$th layer, of the top and bottom slabs
respectively. Since translational invariance is unbroken on the plane,
the eigenstates can be labeled by the momenta ${\bf k}=(k_x, k_y)$.The
two wave-functions of the zeroth layer - $\psi^t_0({\bf k})$,
$\psi^b_0({\bf k})$ - are then matched at the interface in the
presence of the coupling matrix $H_V$ given in
Eq. \ref{eq-coupling}. As shown in detail in Appendix \ref{App:ComputationFermiArc}, the
conditions for the existence of the interface states can be expressed
as a matrix equation
\begin{align}
M(E, \kappa, p_x, p_y) A = 0,
\end{align}
where $M(E, \kappa, p_x, p_y)$ is a (not necessarily hermitian) square
matrix of dimension $2N_{sc}\times2N_{sc}$ and $A$ is a column matrix
with $2N_{sc}$ rows (the factor of 2 is for the spin degrees of
freedom). Here, we have used ${\bf p}=(p_x, p_y)$ to represent the
momenta in the superlattice (SL) BZ.  For interface localised states
to exist, the determinant of the matrix must vanish. Therefore
$\textrm{det}(M(E, \kappa,p_x, p_y))=0$ gives a condition which $(p_x,
p_y)$ must satisfy for a given $E$ and $\kappa$ to have the interface
localised states.The set of all such $(p_x, p_y)$, for $E=0$ and fixed
$\kappa$, yields the reconstructed Fermi arc state at the interface.

\begin{figure*}
\includegraphics[width=0.95\linewidth, height=0.2\linewidth]{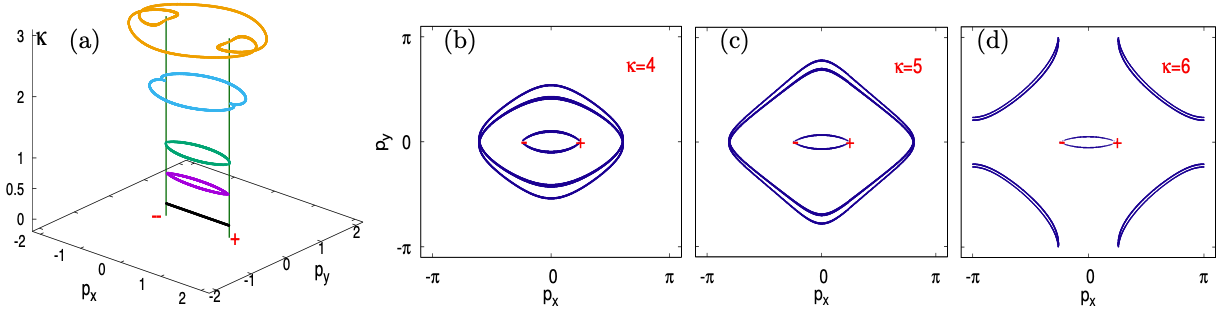}\\
\includegraphics[width=0.95\linewidth, height=0.2\linewidth]{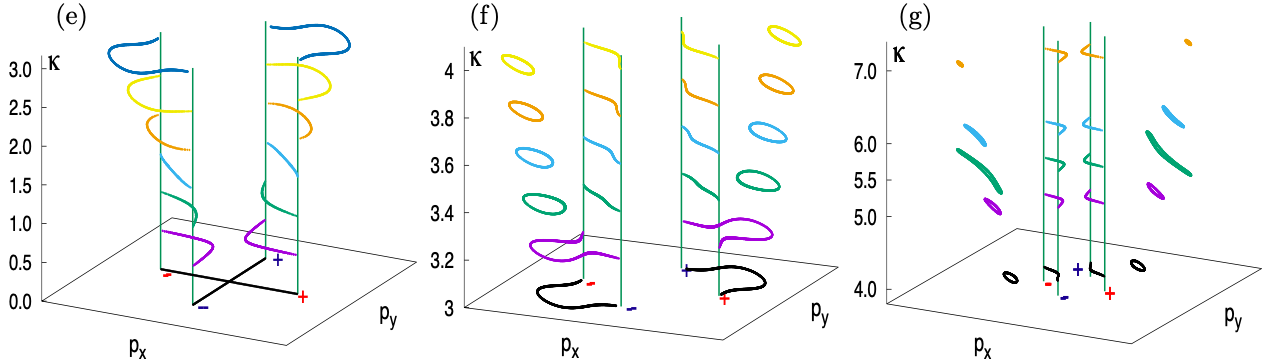}
\caption{The panels show the evolution of the Fermi arcs (FAs) of the interface BZ  as a function of the coupling strength.
The panels in the top row are  for $n_0=0$ and those in the bottom row are for $n_0=1$. (i) Panels (a),(b),(c) and (d)  are for $n_0=0$
and a relative twist between the top and bottom slabs of $2\theta_0=\pi$.   The positive (negative) chiral Weyl point projection  (WPP) of the top slab and the negative (positive) chiral WPP of the bottom slab coincide  at $(+(-)\pi/4, 0)$ in the interface  BZ.  The parameters are $t'=\sin{\pi/3}$,  $k_0=\pi/4$ and $V_x= 1,  V_y=0$. When $\kappa=0$, the FAs of both the slabs are just straight lines along the $x$-axis between $(-\pi/4, \pi/4)$. When we switch on the coupling, the FAs evolve with increasing $\kappa$ as shown in panel  (a). The  vertical green lines represent the WPPs.  For relatively higher values of $\kappa=4, 5, 6$,  the reconstructed FAs are shown separately in panels (b), (c) and (d). (ii) Panels (e), (f), (g) and (h) are for  $n_0=1$ and for different ranges of $\kappa$.
 The parameter values are
the same as those for the top panels. Here, the relative twist between the slabs is  $2\theta_1=\pi/2$. The unreconstructed FA for the bottom slab 
is along the y-axis and for the top slab, the unreconstructed FA is along the  x-axis (shown as black lines in panel (e)).
 As a function of the coupling $\kappa$, the FAs go through a reconstruction which joins the two positive WPPs  and the two negative WPPs 
 of the two slabs together and for some values of $\kappa$ we see Fermi loops disconnected from the WPPs }
\label{fig-n=0-1}
\end{figure*}

To proceed further, we need to fix $n_0$ and explicitly find the FAs
by solving $\textrm{det}(M(E, \kappa,p_x, p_y))=0$. In the next
subsection, we first consider the simplest cases, $n_0=0$ and $n_0=1$,
with respective relative twist angles $2\theta_0=\pi$ and
$2\theta_1=\pi/2$.  In both cases, the two lattices are in registry
and the full $xy$ translation symmetry of the original cubic lattice
is present at the interface. In the following subsection, we will
present results for the $n_0=2$ case in detail, with a larger
interface unit cell, and therefore a reduced translation symmetry. The
results for other twists (which are qualitatively similar) are
presented in Appendix \ref{App:n0=4And5}.

\vspace{0.2cm}

\subsection{Twists with ${\bf n_0=0}$ and  ${\bf n_0=1}$}

\begin{figure*}
\includegraphics[width=0.9\linewidth, height=0.25\linewidth]{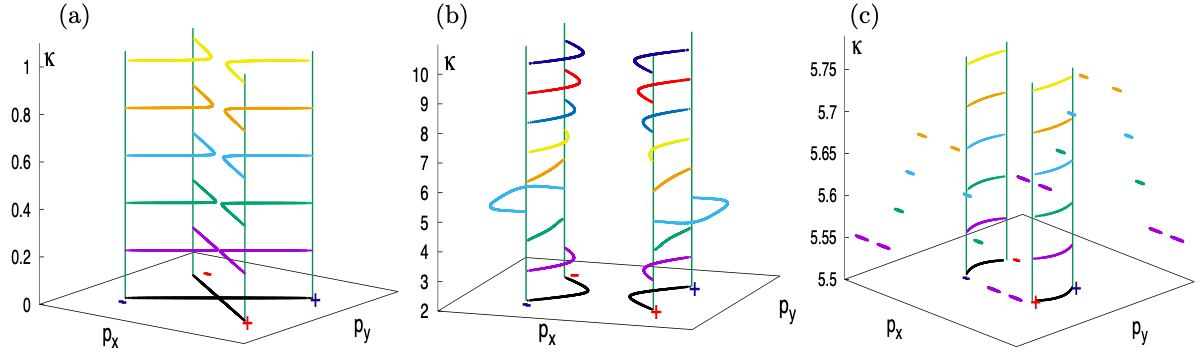}
\includegraphics[width=0.9\linewidth, height=0.25\linewidth]{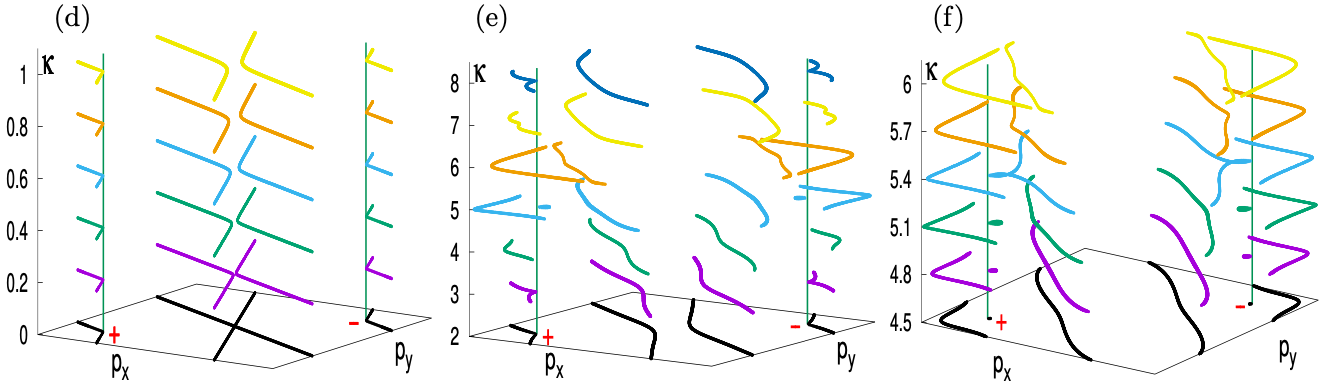}
\caption{The panels show the evolution of the FAs at  the interface BZ  as a function of the coupling strength for 
$n_0=2$. (i) The  panels in the top row show  the evolution in the  non-overlapping case where the Weyl point projections (WPPs) of the top (red $+$ve and $-$ve signs)and bottom slabs (black $+$ve and $-$ve signs)  are distinct from each other. The vertical green lines represent the evolution of the WPPs.  The parameters  chosen are $t'=\sin{\pi/3}$, and $V_x=1, V_y=0$. (a)  Reconstructed FAs are shown for the coupling parameters in the range $\kappa=(0, 1)$. For $\kappa=0$, the  original FAs  of  the individual slabs are shown in black. As $\kappa$ increases, the reconstructed arcs  join the  $+$ve chirality WPPs together  and the $-$ve  chirality WPPs together (b) The range of the coupling is now from $\kappa =2$ to $\kappa=10$.
(c)  Shows the  change of the sign of curvature of the FAs from positive to negative around $\kappa=5$ and formation of pairs of closed Fermi loop at the BZ boundary. These loops disappear pairwise after merging together around $\kappa=5.75$. (ii) The panels in the bottom row show the
evolution when the WPPs of the top and bottom slabs overlap. The parameters  chosen are  $t'=1.5$, $V_x=1$, $V_y =0.2$. 
 (d)  Evolution of the Fermi arcs for small $\kappa$.
(e)  The evolution shows the detachment of the FAs  from the WPPs at around $\kappa \sim 3.5$. Also, the formation of a new small
loop passing through the WPPS is seen around $\kappa \sim 5$. (f)  A close-up view of the formation of the small closed loop passing through 
WPPs  }
\label{fig-n=2}
\end{figure*}

To set the stage for studying arbitrary commensurate angles, we first
study the case $n_0=0$.  This case was studied earlier in Ref.
\onlinecite{Dwivedi2018} for the special case $t'=1$ and the interface
hopping being of the same form as the bulk hopping, except with a
different strength. Here, we consider our more general model, and we
consider the evolution of the FAs as a function of the coupling
parameter. When the slabs are aligned, the Weyl point projections
(WPPs) in the surface BZ of both the top and bottom slabs are at $(\pm
k_0, 0)$ are on top of each other, with the positively charged chiral
WPPs at $(k_0, 0)$ and the negatively charged chiral WPPs at $(-k_0,
0)$.  In this case there need not be any states localized at the
interface. However, when the two slabs are rotated by a relative angle
of $2\theta_0=\pi$, the position of the positively charged chiral WPP
of the top slab coincides, in the surface BZ, with the negatively
charged chiral WPP of the bottom slab.  Since the direction of the FA
has changed between the slabs, the interface is well defined and there
are necessarily interface localized states which get reconstructed as
a function of the tunnel coupling.  Note that the symmetry of the
surface BZ under reflections about the $x$-axis and $y$-axis implies
that the tunnel-couplings $V_x$ and $V_y$ in Eq. \ref{eq:tunneling}
can be chosen to be real and non-zero. For definiteness, we have
chosen $V_x \ne 0$, although we have checked that taking a general
linear combination of $V_x$ and $V_y$ does not change the result
qualitatively.  As mentioned above, the reconstructed FAs for $E=0$
and for a fixed coupling parameter $\kappa$ is given by the set of
momenta $(p_x, p_y)$ for which the determinant of the $4\times4$
matrix $M(E, \kappa, p_x, p_y)$ vanishes.  The results are shown in
the panels in the top row of Fig. \ref{fig-n=0-1} for a set of values
of $\kappa$. At $\kappa=0$, the FA of the individual slabs are just
straight lines on the $x$-axis between $(-k_0, k_0)$.  With the
increase in the coupling parameter $\kappa$, the FAs evolve as shown
in Fig. \ref{fig-n=0-1}(a). At $\kappa \sim 4$, in addition to the
main FA connecting the WPPs, a pair of freestanding Fermi loops
appear.  For larger $\kappa$, the freestanding loops disappear and
eventually, beyond the values shown in the graph, we get back the
result for the FAs of the decoupled slabs.  This is a particular
instance of the duality in Fermi arc reconstruction under
$\kappa\rightarrow 1/\kappa$ which is seen in all cases and which we
shall discuss in detail later.

Next, let us consider the case when $n_0=1$.  This was also earlier
studied in Ref. \onlinecite{Murthy2020}, but only for weak values of
the coupling $\kappa$. When the slabs are rotated with respect to each
other by an angle $2\theta_1=\pi/2$, the WPPs of one of the slabs in
the surface BZ is $(0, \pm k_0)$, whereas those of the other one are
at $(\pm k_0, 0)$. As for $n_0=0$, the 
lattices are in registry and the reconstructed FAs are given by
by a set of momenta $(p_x, p_y)$ for which the determinant of the
$4\times 4$ matrix $M(E, \kappa, p_x, p_y)$ vanishes for a given
$\kappa$ and for $E=0$.  The results for the reconstructed FAs are
shown in the bottom row of panels in Fig. \ref{fig-n=0-1}.  As
expected, the FAs at zero coupling $\kappa$ are the FAs of the
individual slabs. For the upper slab, it is a straight line on the
$x$-axis and for the lower slab, it is a straight line on the
$y$-axis. For nonzero $\kappa$, the reconstructed FAs connect the
positively charged chiral WPPs and the negatively charged chiral WPPs
of the slabs together as shown in Fig. \ref{fig-n=0-1}(e). As $\kappa$
increases further, the FAs get deformed and a pair of freestanding
closed loops form. For even larger $\kappa$, the freestanding loops
disappear and eventually, as in the earlier case, the reconstructed
FAs approaches the zero-coupling result.

{\subsection{Twist with ${\bf n_0=2}$}}

The simplest case where the surfaces of the top and bottom slabs are
not in registry occurs for $n_0= 2$, where the top and bottom slabs
are twisted clockwise and counter-clockwise respectively by the angle
$\theta_2=\tan^{-1}(1/2)$. This brings the lattice point $(2, 1)$ of
the bottom layer of the top slab lie on top of the $(2, -1)$ site of
the top layer of the bottom slab, as shown in Fig. \ref{fig-SL},
forming one of the sites of the superlattice (SL). As can be seen from
the figure, the SL unit cell contains 5 lattice points from each of
the layers ($N_{sc}=2\times(n_0^2 +1)=10$ lattice sites altogether),
and its unit cell is 5 times as large as the original unit cell.  The
primitive lattice vectors of the rotated slabs are now given by ${\bf
  a}_{t1}=(\cos{\theta_2}, -\sin{\theta_2})$, ${\bf
  a}_{t2}=(\sin{\theta_2}, \cos{\theta_2})$ for the top layer and
${\bf a}_{b1}=(\cos{\theta_2}, \sin{\theta_2})$, ${\bf
  a}_{b2}=(-\sin{\theta_2}, \cos{\theta_2})$ for the bottom layer.
The superlattice primitive vectors are given by ${\bf b}_1=(\sqrt{5},
0)$ and ${\bf b}_2=(0, \sqrt{5})$ and the SL unit cell contain total
$N_{sc}= 2(n_0^2 + 1) = 10$ lattice sites as shown in
Fig. \ref{fig-SL}. The WPPs in the surface BZ of top slab are given by
(i) $`+' $ at $k_0(\cos{\theta_2}, -\sin{\theta_2})$ and $`-'$ at
$k_0(-\cos{\theta_2}, \sin{\theta_2})$ and that of bottom slab are
given by (ii) $`+'$ at $ k_0(\cos{\theta_2}, \sin{\theta_2})$ and
$`-'$ at $ k_0(-\cos{\theta_2}, -\sin{\theta_2})$.  Note that our
model does not specify $k_0$ - see Eq. \ref{eq:ham} - it only says that
the Weyl nodes are at $(\pm k_0, 0,0)$ - hence we are free to choose
it to be any value consistent with the existence of FA states. Without
loss of generality, we choose an arbitrary value of $k_0=\pi/4$.

The results for this case are shown in Fig. \ref{fig-n=2}.  As we turn
on the coupling parameter $\kappa$, the reconstructed FAs connect the
$+$ve chiral WPPs together (and the $-$ve chiral WPPs together). As
$\kappa$ is increased, the curvature of the reconstructed FAs changes
and flips sign near $\kappa \sim 5$ as shown in
Fig. \ref{fig-n=2}(b). This range of coupling $\kappa=(5.5, 5.8)$ is
explored further in Fig. \ref{fig-n=2}(c), where a set of four small,
closed, freestanding Fermi loops appear at the corners of the BZ when
$\kappa \sim 5.55$. When $\kappa$ is further increased, they move
towards each other, merge, and finally disappear at $\kappa \sim
5.75$.  As for  $n_0=0,1$, there is a duality in the FA
reconstruction between small and large $\kappa$; at large $\kappa$ we
get qualitatively the same FAs as small $\kappa$. In
Fig. \ref{fig-n=2}(b), one can see that the FAs at $\kappa=2, 3$ are
similar to the FAs at $\kappa=10, 9$.  Note that, in this case, we see
that the reconstructed Fermi arcs are always attached to the WPPs.

Ref. \onlinecite{Murthy2020} had conjectured that there might be
arcless angles in twisted WSMs, where all interface states are
disconnected from the WPPs. Can that occur in our model with
commensurate twist angles as well? A precondition for this is to have
WPPs of the {\it same} chirality from the top and bottom slabs overlap
in the surface BZ. This can be achieved by combining a commensurate
twist angle with a suitable choice of $k_0$.  For example, at $n_0=2$,
we rotate the top slab clockwise by an angle $\theta_2=\tan^{-1}(1/2)$
and the bottom slab by the angle $(\pi/2 + \theta_2)$ anti
clockwise. The SL at the interface is identical to the earlier case,
but the primitive lattice vectors are now given by ${\bf
  a}_{t1}=(\cos{\theta_2}, -\sin{\theta_2})$, ${\bf
  a}_{t2}=(\sin{\theta_2}, \cos{\theta_2})$ for the top slab and ${\bf
  a}_{b1}=(-\sin{\theta_2}, \cos{\theta_2})$, ${\bf
  a}_{b2}=(-\cos{\theta_2}, -\sin{\theta_2})$ for the bottom slab. Now
we can choose $k_0=2\pi/3$ so that the positively charged chiral WPP
of both the slabs coincide (and similarly the negatively charged
chiral WPPs). The positively and negatively charged chiral WPPs are
then at $k_0 (\mp1/\sqrt{5}, \mp 1/\sqrt{5})$ in the surface BZ.

The reconstructed FAs in this case are shown in the panels in the
bottom row of Fig. \ref{fig-n=2}. Fig. \ref{fig-n=2}(d) shows the
situation for weak-coupling, where the FAs are attached to the WPPs.
Fig. \ref{fig-n=2}(e) reveals that FAs get detached from the WPPs for
$\kappa \sim 3.5 $ and move away from them as $\kappa$
increases. Subsequently a pair of small closed Fermi loops attached to
the WPPs appear at $\kappa \sim 4.5$. Upon increasing $\kappa$ they
disappear at $\kappa \sim 5.5$ (see Fig. \ref{fig-n=2}(f)). So in the
range of $\kappa$ between $3.5<\kappa<4.5$, and $5.5<\kappa<7.5$, the
model has FAs which are wholly disconnected from the WPPs and there
are no surface states attached to the WPPs.  Beyond $\kappa = 7.5$, the results are
similar to the case when the couplings are small, because of the
duality between large and small couplings.

We have also studied the twists with $n_0=4$ and $n_0=5$ for
completeness. Since the results are qualitatively similar to the cases
with smaller $n_0$, they are relegated to  Appendix \ref{App:n0=4And5}.

\textcolor{magenta}{\section{Model for a lift-off transition}}

  We have seen in the previous section that when the projections of
  the Weyl points of positive chirality coincide in the surface
  Brillouin zone (and likewise for the negative chirality Weyl
  points), the Fermi arc can detach from the WPP at an appropriate
  coupling strength $\kappa$. Because of the $\kappa\rightarrow
  1/\kappa$ duality, the Fermi arc re-attaches again to the WPPs when
  $\kappa$ is increased and the reconstructed Fermi arc eventually
  approaches the unreconstructed Fermi arc in the limit $\kappa^{-1}
  \to 0$. In this section, our goal is to demonstrate these lift-off
  and re-attachment transitions in a simple model where there is no
  twist between the slabs. More specifically, we want to obtain the
  shape of the segment of the Fermi arc attached to the WPP at the
  transition, which, based on empirical evidence, we believe to be
  universal.

\begin{figure*}[ht]
\includegraphics[width=0.8\linewidth, height=0.3\linewidth]{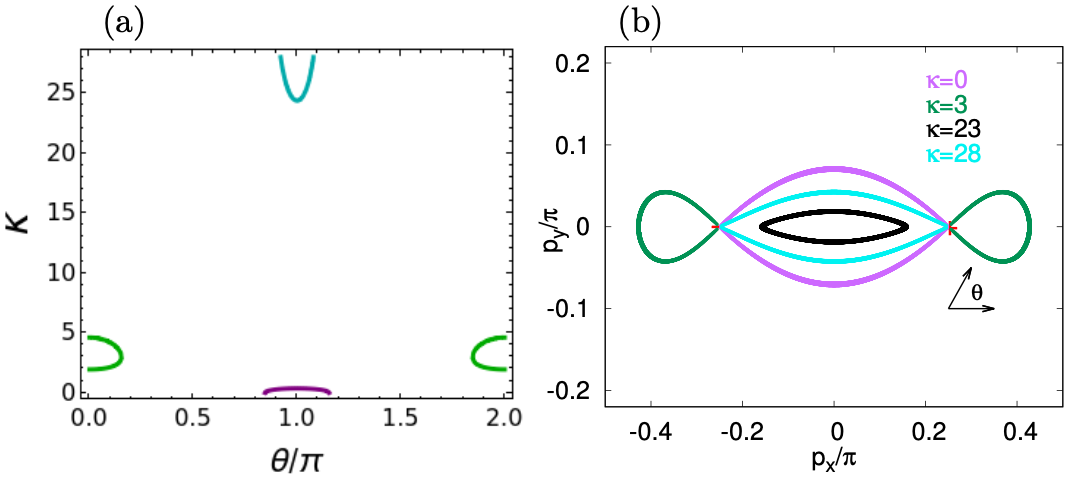}
\caption{The lift-off transition (a) The curves denote the values of the coupling $\kappa$ as
  a function of $\theta$ for the existence of Fermi arcs in the
  vicinity of the Weyl point projection (WPP) at ${\bf k} =(\pi/4, 0)$
  and with $V=\sigma_x + it'\sigma_y$.  The parameters chosen are
  $\lambda=0.75$ and $t'=0.8$.  The solutions around $\theta=\pi$
  correspond to the Fermi arcs which are in purple($\kappa=0$) and
  cyan($\kappa=28$) in figure (b), whereas the solutions around
  $\theta=0$(or $2\pi$) corresponds to the Fermi arcs in green. There
  are two $\kappa$ regions ( $0.37<\kappa<1.95$ and
  $4.62<\kappa<24.35$ ) where there are no Fermi arcs attached to the
  WPPs. As $\kappa$ is increased slowly from zero, the Fermi arc in
  purple colour in figure (b) evolves until it detaches from the WPPs
  at $\kappa_{c_1}\approx0.37$. Then in the
  parameter regime, $0.37<\kappa<1.95$, there are no solutions until
  $\kappa_{c_2}\approx1.95$. In the regime, $1.95<\kappa<4.62$, the
  Fermi arc in green evolves and detaches from the WPPs at the
  critical value $\kappa_{c_3}\approx4.62$.  Once again , there are no
  solutions when $4.62<\kappa<24.35$. Beyond $\kappa>24.35$ we get the
  dual solution, the Fermi arc in cyan in figure (b) which gradually
  approaches the zero coupling Fermi arc (in purple) as $\kappa \to
  \infty$.  }
\label{fig-liftoff}
\end{figure*}

We consider the Hamiltonian given in Eq. \ref{eq-wsm0} in the main
text, but with a modified top/bottom-dependent $f_y$ term:
\begin{align*}
f_{y,\gamma}= \sin{k_y} + \lambda_{\gamma} (\cos{k_x} - \cos{k_0}).
\end{align*}
For the top slab ($\gamma \equiv t$) we take $\lambda_t = \lambda$ and
for the bottom slab ($\gamma\equiv b$), $\lambda_b = -\lambda $. There
is no twist, but the slabs, though aligned, are not identical even in
the bulk except at $\lambda=0$. We consider short-range hopping as
before (Eq. \ref{eq-V-short}) and the hopping matrix $V$ is taken to be
$V = \sigma_x + i t' \sigma_y $.

As mentioned before, we need to compute the determinant of the matrix
$M(E, \kappa, k_x, k_y)$ which is now a $4 \times 4$ matrix, in the
neighbourhood of the projection of the Weyl point.  So we will
parametrize the neighbourhood of the Weyl point projection ${\bf k} =
(k_0, 0)$ in the surface BZ as
\begin{align}
\vec{ k}=k_0 \hat{x} +  \vec{q}, \hspace{0.5cm}  \vec{q} = q (\cos{\theta} \hat{x} +  \sin{\theta} \hat{y}),
\end{align} 
where $q = |\vec{q}|\ll 1$ and $\theta $ is the polar angle in the
two-dimensional surface BZ around the Weyl point. The matrix $M(E,
\kappa, k_x, k_y)$ can be explicitly written as
\begin{align}\label{eq-liftoffmatrix}
M= \begin{pmatrix} % or pmatrix or bmatrix or Bmatrix or ...
      \frac{g_{1\uparrow}^{(t)}}{u_1^{(t)}}& \frac{g_{2\uparrow}^{(t)} }{u_2^{(t)}} & -\kappa g_{1\uparrow}^{(b)} &  -\kappa g_{2\uparrow}^{(b)} \\
      \frac{g_{1\downarrow}^{(t)}}{ u_1^{(t)}} & \frac{g_{2\downarrow}^{(t)}}{u_2^{(t)}} & -\kappa g_{1\downarrow}^{(b)} &  -\kappa g_{2\downarrow}^{(b)} \\
      \kappa g_{1\uparrow}^{(t)} & \kappa g_{2\uparrow}^{(t)}  & - \frac{g_{1\uparrow}^{(b)}}{ u_1^{(b)}} & -\frac{g_{2\uparrow}^{(b)}}{ u_2^{(b)}}  \\
      \kappa g_{1\downarrow}^{(t)} &  \kappa g_{2\downarrow}^{(t)} &  -\frac{g_{1\downarrow}^{(b)}}{u_1^{(b)}} & -\frac{g_{2\downarrow}^{(b)}}{u_2^{(b)}}
   \end{pmatrix}
\end{align}
where $u^{(t)}$ and $u^{(b)}$ are determined by
Eqs. \ref{eq-u}-\ref{eq-xi} and the various $g$'s are the components
of the spin wavefunction of the top slab $\Phi^t =
(g_{\uparrow}^{(t)}, g_{ \downarrow}^{(t)})^T$ and the bottom slab
$\Phi^b = (g_{\uparrow}^{(b)}, g_{ \downarrow}^{(b)})^T$. The
arguments of the matrix elements have been suppressed for notational
simplicity. Our strategy is to Taylor expand each $u$ and $g$ to
leading order in q. After some algebra, we get the following leading
order expansions for the u's :
\begin{align}\label{eq-lift_u}
u^{(t)}_1 &= 1 - \frac{q}{t'} s_{t}(\theta),  \hspace{0.5cm}  u^{(b)}_1 = 1 - \frac{q}{t'} s_b(\theta),  \nonumber \\
u^{(t)}_2 &= u^{(b)}_2 = \frac{1-t'}{1+t'} (1 + \sin{k_0} ~ q\cos{\theta} ),  \end{align}
where $s^2_{\gamma}(\theta)  =  (1+\lambda^2) \sin^2{k_0} \cos^2{\theta} - \lambda_{\gamma} \sin{k_0}\sin{2\theta} + \sin^2{\theta}$.

To avoid singularities, we now divide the whole $\theta$ region into
two sub regions - (i) $\pi/2<\theta<3\pi/2$ and (ii) $ -\pi/2<\theta
<\pi/2$ - and choose the spinors appropriately.  The leading order
expansions give the following solutions for the $g$'s - \\ In region
(i) :
  \begin{align}
g^{(t)}_{1\downarrow} & =  \frac{\sin{\theta} - \lambda \sin{k_0} \cos{\theta}}{-\sin{k_0} \cos{\theta}  + s_{t}(\theta)}, \hspace{0.5cm} g^{(t)}_{2\downarrow}= \mathcal{O}(q^2) \nonumber \\
g^{(b)}_{1\uparrow} & =  -\frac{\sin{\theta} + \lambda \sin{k_0} \cos{\theta}}{-\sin{k_0} \cos{\theta}  + s_{b}(\theta)}, \hspace{0.3cm} g^{(b)}_{2\uparrow}= \mathcal{O}(q^2)  
\end{align}
where we have chosen $g^{(t)}_{1\uparrow}=g^{(t)}_{2\uparrow}=1$, $g^{(b)}_{1\downarrow}=g^{(b)}_{2\downarrow}=1$. \\
In region (ii): \\
\begin{align}
g^{(t)}_{1\uparrow} & =  \frac{\sin{\theta} - \lambda \sin{k_0} \cos{\theta}}{\sin{k_0} \cos{\theta}  + s_{t}(\theta)}, \hspace{0.5cm} g^{(t)}_{2\downarrow}= \mathcal{O}(q^2) \nonumber \\
g^{(b)}_{1\downarrow} & =  -\frac{\sin{\theta} + \lambda \sin{k_0} \cos{\theta}}{\sin{k_0} \cos{\theta}  + s_{b}(\theta)}, \hspace{0.3cm} g^{(b)}_{2\uparrow}= \mathcal{O}(q^2) 
\end{align}
where we have chosen $g^{(t)}_{1\downarrow}=g^{(t)}_{2\uparrow}=1$,
$g^{(b)}_{1\uparrow}=g^{(b)}_{2\downarrow}=1$.

We are interested in the existence of Fermi arcs in the vicinity of
Weyl point projection. So for $q \ll 1$, we can approximate $u_1^{(t)}
= u_1^{(b)} \approx 1$ and $u_2^{(t)} = u_2^{(b)} = u_2 \approx
\frac{1-t'}{1+t'}$ (see Eq. \ref{eq-lift_u}).  This essentially
implies that, very close to the Weyl point, the polar angle of ${\bf
  q}$, or the angle at which the FA attaches to the WPP, is the
important parameter.  Substituting these expressions in the matrix
$M$, we get the following determinant vanishing conditions in the two
regions -\\ For region (i) $\pi/2<\theta<3\pi/2$:
\begin{align}\label{eq-lift_theta1}
 (\alpha_t \alpha_b) \kappa^4  + \frac{\kappa^2}{u_2^2}\left((1 + u_2^2) -  2u_2(1+ \alpha_t \alpha_b) \right)  +\frac{\alpha_t \alpha_b}{u_2^2} = 0 , 
\end{align}
and for  region (ii) $ -\pi/2<\theta <\pi/2$:
\begin{align}\label{eq-lift_theta2}
  \kappa^4  + \frac{\kappa^2}{u_2^2}\left((1 + u_2^2) \alpha_t \alpha_b  -  2u_2(1+ \alpha_t \alpha_b) \right)  +\frac{1}{u_2^2} = 0 , 
\end{align}
where $\alpha_t = (\sin{\theta} - \lambda \sin{k_0}
\cos{\theta})/(\sin{k_0} |\cos{\theta}| + s_{t}(\theta))$ and
$\alpha_b = (\sin{\theta} + \lambda \sin{k_0} \cos{\theta})/(\sin{k_0}
|\cos{\theta}| + s_{b}(\theta))$.  The solutions to
Eqs.\ref{eq-lift_theta1}-\ref{eq-lift_theta2} for $\kappa$ as a
function of $\theta$ give us the Fermi arcs attached to the WPP's.
The curves covered by these solutions are shown in
Fig. \ref{fig-liftoff}(a) and in the remaining region, not covered by
these lines, there are no Fermi arcs attached to the
WPPs. Essentially, if there exist surface states in the remaining
regions, those solutions are not coupled to the WPPs and would
actually form closed Fermi surfaces.  However, we do not study such
solutions here since our aim here was to study the lift-off transition
or the limits in the $(\kappa,\theta)$ space, where Fermi arcs
attached to the WPPs are no longer present.

\begin{figure}[ht]
\includegraphics[width=1\linewidth, height=0.45\linewidth]{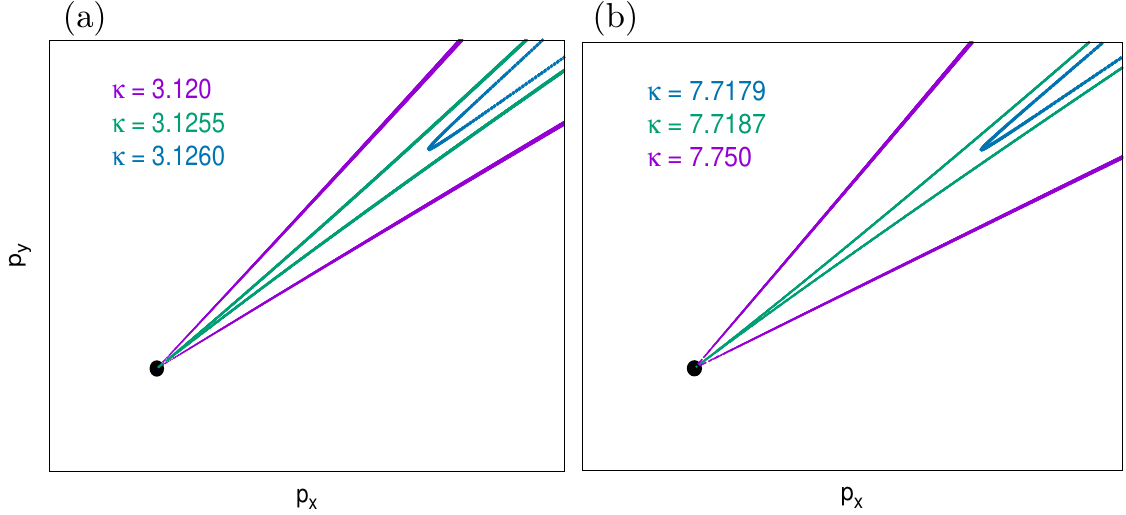}
\caption{Singular shape of the Fermi arc near the (a) lift-off and (b)
  re-attachment transitions for $n_0=2$ (see also the reconstructed
  Fermi arc in Fig. \ref{fig-n=2}(e)). The black dot in the figure
  represents the projection of negative chirality Weyl node.  }
\label{fig-n2shape}
\end{figure}

We notice that at and near the lift-off (or re-attachment) transition,
the shape of the Fermi arc is highly singular. The two legs that are
attached to the WPP have the same slope at the transition. We believe
this shape to be universal for all lift-off and re-attachment
transitions, because in every instance of lift-off or re-attachment
where we have ``zoomed in'' near the WPP at the transition, we find this
to be the case. An example is shown in Fig. \ref{fig-n2shape}.

\textcolor{magenta}{ \section{Caveats, conclusions, discussion, and open questions}}
 \label{sec:dc}

 In this work we considered two identical semi-infinite slabs of WSM,
 twisted by a commensurate angle with respect to each other, and their
 free surfaces tunnel-coupled. The constraint that the angle be
 commensurate means that a reduced lattice translation symmetry,
 defined by a larger superlattice unit cell, is enjoyed by the
 Hamiltonian. This has the benefit of allowing us to extend our
 calculations to arbitrary values of the tunnel-coupling, which
 allowed us to extend the previous results of Murthy, Shimshoni, and
 Fertig \cite{Murthy2020} (which were for arbitrary small (incommensurate) angles, but
 perturbative values of the tunnel couplings.)

 At this point it is important to mention several caveats concerning
 the assumptions we have made. Firstly, we focused most of our
 investigation on the case when the tunnel-couplings are
 ultra-short-range, being nonzero only for sites on the two surfaces
 having the same $xy$ coordinates. We did investigate longer-range
 hoppings as perturbations to this case (Appendix
 \ref{App:LongRangeHopping}), but did not undertake a study of the
 most general periodic hopping. Secondly, even with this
 simplification, the number of parameters in the hopping matrix is too
 large to allow systematic investigation. We therefore chose a
 particular subset of symmetries of the model system in the absence of
 tunnel-coupling, and imposed this symmetry on the tunnel-couplings as
 well. Once again, we have checked perturbatively that adding
 tunnel-coupling that break our self-imposed symmetries do not change
 our results qualitatively (Appendix \ref{App:SymmetryPerturbation}).
 
 Three main results emerge from our work. Firstly, we confirmed an
 interesting conjecture from earlier work. MSF \cite{Murthy2020}
 considered (among other things) the case when the twist angle and the
 value of $k_0$ are tuned such that the WPPs of the $+$ Weyl nodes of
 the two slabs overlap (and likewise for the WPPs of the $-$ Weyl
 nodes).  MSF conjectured that at strong enough tunnel-couplings the
 reconstructed Fermi arcs would detach themselves from the WPPs,
 leaving behind purely interface states, that is, states with all
 their spectral weight near the interface. We have confirmed that this
 appears to be a generic feature when the WPPs overlap. In addition,
 free-floating Fermi loops far from any WPP appear, expand, contract,
 and disappear as a function of the strength of the tunnel-coupling.

 A noteworthy feature of such purely interface states (which they
 share with the surface states of topological insulators) is that some
 of them are two-dimensional states that cannot be obtained in a
 purely two-dimensional system of noninteracting electrons. For
 example, consider Fig. \ref{fig-n4-detach}a, with $n_0=4$ and
 $\kappa=5.2$. The entire Fermi loop is one connected curve winding
 around the SL BZ (periodic boundary conditions apply at the
 boundaries of the SL BZ). However, it has no inside or outside. There
 is no notion of the number of states enclosed inside the Fermi
 surface.
 
 Secondly, we identified a qualitative duality between weak and strong
 tunnel-couplings. This occurs for a very physical reason. Let us
 restrict ourselves to the case when the tunnel-couplings are
 ``zero-range'', in the sense that the lattice sites of the two layers
 have to have identical $xy$ coordinates in order for the hopping to
 be nonzero. A very strong tunnel coupling $\kappa$ between the two
 vertically aligned sites, considered in isolation, will create a pair
 of hybridized states of energy of order $\pm\kappa$. Any tunneling
 between the slabs must go through these vertically aligned
 sites. Thus, the effective tunneling must be of order $t^2/|\kappa|$,
 where $t$ is the intra-slab tunneling strength. This shows that the
 two slabs become essentially isolated from each other as
 $|\kappa|\to\infty$. It may be possible to engineer large values of
 $|\kappa|$ in WSMs that cleave such that atoms on one WSM surface are
 likely to form covalent bonds with atoms on the other WSM surface.

 Thirdly, we looked at the shape of the reconstructed Fermi arcs at
 the lift-off and re-attachment transitions. We found that they have a
 very singular shape, as demonstrated in Fig. \ref{fig-n2shape}. The
 singularity near the WPP also seems to be universal, in the sense
 that all lift-off and re-attachment transitions we have investigated
 in detail show the same shape near the WPP. 

 Let us examine potential experimental signatures of our theoretical
 results. First we consider the closed Fermi curves of purely
 two-dimensional interface states, as exemplified by Fig. \ref{fig-n4-detach}a. Upon
 applying a weak (semiclassical) perpendicular orbital magnetic field,
 wave packets will experience a Lorentz force $-e{\bf v}({\bf
   k})\times{\bf B}$ where ${\bf v}({\bf k})={\bf\nabla}_{\bf
   k}\varepsilon({\bf k})$ is the group velocity of the Fermi loop
 states. Since the velocity is always perpendicular to the Fermi
 loop, the wave packets semiclassically travel along the closed Fermi
 curves. Under semiclassical quantization the states along the Fermi
 curve re-organize themselves into a set of equally spaced levels,
 with the the spacing directly proportional to $|{\bf B}|$. These
 levels can be investigated by the absorption of electromagnetic waves
 of the appropriate frequency.

 How might one detect lift-off/reattachment transitions? A standard
 technique \cite{Potter2014} to look for Fermi arc states passing
 through the WPPs is to look at semiclassical orbits (once again under
 a weak perpendicular magnetic field) that traverse the Fermi arc on
 one surface, go through the bulk via the Weyl node to the other
 surface, traverse the Fermi arc there and complete the cycle.Such
 intersurface loops can enclose an area, and exhibit
 magneto-oscillations. \cite{Moll2016} The period of the cycle depends
 on the thickness of the slab. As usual, semiclassical quantization
 will reorganize the closed orbits into a set of equally spaced level,
 which can be investigated by an electromagnetic probe.

 To be more specific on the experimental signature of the
 liftoff/re-attachment transitions, let us focus on the case when the
 WPPs overlap, and we are at weak-coupling, such that the Fermi arcs
 are attached to the WPPs. A wavepacket starting on the bottom surface
 of the lower slab will traverse the bulk of the lower slab through
 the Weyl node and reach the interface of the two slabs. At this point
 it will split; a part will traverse the Fermi arc at the interface,
 and another part will travel through the bulk of the upper slab,
 traverse the Fermi arc on the top surface of the upper slab, and
 return to the interface. Thus, there will be multiple scattering of
 the wavepacket, giving rise to a sequence of periods of return of the
 wavepacket to the bottom surface of the lower slab. Similarly, there
 will be a sequence of areas relevant to magneto-oscillations.

 Now, if the tunnel-coupling is tuned such that the Fermi arcs detach
 from the WPPs, the interface is inaccessible to a wavepacket starting
 on the bottom surface of the lower slab. Thus, there is only one
 period of return for the wavepacket and only one area relevant to
 magneto-oscillations.

Thus, if an {\it in-situ} method (perhaps pressure) can be found to
tune the strength of the tunnel-coupling through the
lift-off/re-attachment transition, this abrupt change in behavior of
the return period and/or magneto-oscillations will be a smoking-gun
signature of such a transition.

The most important physics left out of our calculation is the effect
of disorder. With regard to disorder, despite early work indicating
the stability of the Weyl node against weak disorder,
\cite{WSM+Disorder-early} a consensus has emerged that large rare
regions of strong disorder potential produce a nonzero density of
states at the Weyl points, destroying the WSM even for arbitrarily
weak disorder. \cite{WSM+Disorder-Global} Similarly, the Fermi arcs get
broadened by coupling to bulk disorder, and conduct dissipatively
 \cite{WSM+Disorder-dissipative-FA} on short to intermediate length
scales. They get localized at the longest length scales, but the
chiral velocity persists at the surface. \cite{WSM+Disorder-FA} Based
on this picture, we can surmise that the Fermi loops traversing the
superlattice BZ that we find in our work should be detectable as conducting
states at all but the longest length scales. However, the states near
the WPPs at the liftoff/re-attachment transitions will be particularly
susceptible to disorder, and may be harder to detect via conduction.

There are many other interesting open questions, such as the
possibility of interface quantum Hall effects and electron-electron
interactions, which we hope to study in future work.

 \vspace{0.2cm}

%--------------------------------------------------------------------------------------------------------------------------------------------------------------
%--------------------------------------------------------------------------------------------------------------------------------------------------------------
 \acknowledgements
 
 SR and GM would like to thank the VAJRA scheme of SERB, India for its
 support.  GM is grateful for partial support from the US-Israel
 Binational Science Foundation (Grant No. 2016130), and the
 hospitality of the International Center for Theoretical Sciences,
 Bangalore, where these ideas were conceived during the workshop on
 Edge Dynamics in Topological Phases, Dec 2019 - Jan 2020.

\vspace{0.5cm}
%--------------------------------------------------------------------------------------------------------------------------------------------------------------- 
\appendix

\centerline{\large \textcolor{magenta}{APPENDICES}}
\vspace{0.5cm}

The appendices provide details on: (a) The stability of FA
reconstructions under longer ranged hoppings.  (b) The symmetries of
the $\kappa=0$ Hamiltonian in the superlattice Brillouin zone and the
implications for tunnel-couplings. (c) The stability of FA
reconstructions under perturbations of tunnelling matrix that break
the above symmetries. (d) The details of the computation of the Fermi
arcs at the interface. (e) Some illustrative results for larger values
of $n_0$, or smaller twist angles.

%---------------------------------------------------------------------------------------------------------------------------------------------------------------
%\textcolor{magenta}{\section{Stability under longer ranged hoppings}} 
\section{Stability of Fermi arc reconstructions under longer ranged hoppings}
\label{App:LongRangeHopping}

\begin{figure*}[ht]
\includegraphics[width=0.9\linewidth, height=0.25\linewidth]{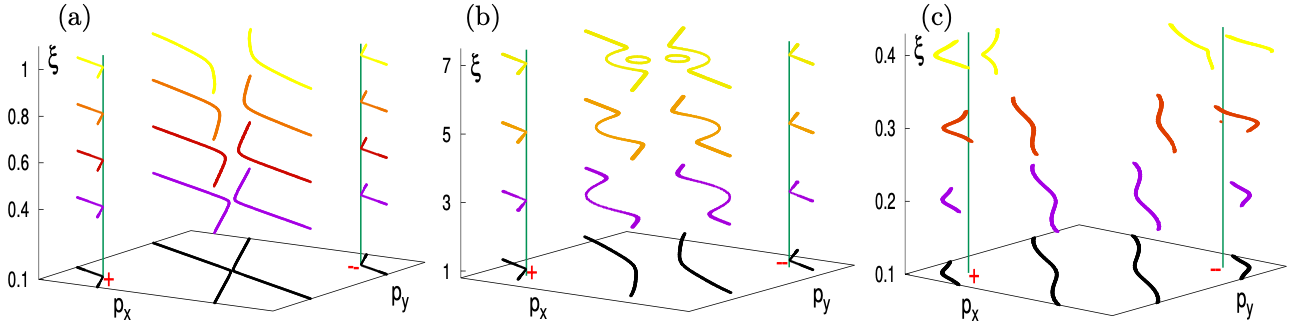}
\caption{$n_0=2$:  Long ranged couplings. The evolution of reconstructed Fermi arcs as a function of the  coupling range parameter $\xi$ is shown for different fixed coupling strengths $\kappa$. The parameters are $k_0=2\pi/3, t'=1.5$,  $V_x=1, V_y=0$. (a) and (b) show the evolution of the reconstructed Fermi arcs with $\xi$ for  $\kappa=0.2$. When $\xi$ is small we recover our previous short-range coupling results(see Fig. \ref{fig-n=2}d)  and it is clear that  the results are stable with the increase in the range of the  couplings. (c) Here, the coupling strength is fixed at $\kappa=4$ to examine  the stability of the Fermi arc detachment from WPPs. $\xi$  is kept small because the effective strength of the coupling  gets renormalized to larger value with increasing $\xi$.  Note that  in (c),  the reconstructed Fermi arcs around  `+' WPP   and   `-'  WPP  are not symmetric for larger values of $\xi$. Indeed,  we do not expect the symmetry 
between the Weyl node projections to exist  for a general $\xi$.     }
\label{fig-n2-longrange}
\end{figure*}

So far we have discussed results where the hoppings between the top
and bottom layers are taken to be ultra-short range, such that only
sites of the top and bottom slabs with the same 2D coordinates are
tunnel-coupled. It is natural to ask what happens to the Fermi arc
states at the interface if the range of the hoppings were increased to
also include the next nearest sites and the next-next nearest sites
and so on. We attempt to answer this question here by considering the
hopping $V_{ss'}({|{\bf r}_t - {\bf r}_b|})$ as a Gaussian,  
\begin{align}
V_{ss'}(|{\bf r}_t - {\bf r}_b |) = \kappa (V_0)_{ss'} e^{-\frac{r^2}{\xi^2}}.
\end{align}
Here $(V_0)_{ss'}$ is a constant matrix, $\kappa$ denotes the strength
of the interaction and $\xi$ is a length scale parameter that
determines the range of hopping. Small(large) $\xi$ means that the
hopping is short(long) ranged. For $\xi<<1$, we should recover our
previous results, which was for ultra-short range hopping.  Without
loss of generality, we study the case $n_0=2$.  In particular, we
consider the case with overlapping Weyl point projections(WPPs).

The results are shown in Fig. \ref{fig-n2-longrange}. We have studied
how the reconstructed Fermi arcs get modified as a function of the
hopping range parameter $\xi$. We have considered two different
coupling strengths $\kappa=0.2$ and $\kappa=4$. Since the effective
coupling strength parameter $\kappa e^{-{r^2}/{\xi^2} }$ gets
renormalized to larger value with increasing hopping range $\xi$, we
restrict the hopping range parameter to smaller values for the latter
case $\kappa=4$. We recover our previous results of short range
hopping for very small $\xi<<1$. We find that the short-range hopping
result is stable against longer ranged hoppings unless $\xi$ is too
large.

%----------------------------------------------------------------------------------------------------------------------------------------------------------
%\textcolor{magenta}{\section{Symmetries of the superlattice Brillouin zone}}
\section{Symmetries of the $\kappa=0$ Hamiltonian in the superlattice Brillouin zone}
\label{App:Symmetry}

In the main text, we have mentioned that the symmetries of the
Hamiltonian of the slabs ($H^t + H^b$) can be kept intact if we
restrict the tunnelling matrix to be of the following form $V =
V_{x}\sigma_x + iV_{y}\sigma_y$, where $V_{i}$ ($i=x, y$) is a real
number.  We will show this result here explicitly for a particular
value of the twist angle $\theta_{n_0=2}$; however, the result is
general and is valid for all $n_0$ .

We will consider the case $n_0=2$ with overlapping Weyl point projections (WPPs) i.e. for commensurate twist angle of the form $\theta_{n_0} = \pi/2 + \tan^{-1}(1/n_0)$. The resulting superlattice (SL) is identical to what is shown in Fig. \ref{fig-SL} (but now with  differently oriented lattice vectors ${\bf a}^{\gamma}_i$, where  $\gamma=t, b$  and  $i=1, 2$).  To analyse the symmetries, it is convenient to take the Hamiltonian (in Eq. \ref{eq-slabHamiltonian}) in the position space in all  directions -
\begin{align}
H^{\gamma}  =  \sum_{{\bf n}} & \hspace{0.3cm}  2(2+\cos{k_0}) c^{\dagger}_s({\bf n})(\sigma_x)_{ss'} c_{s'}({\bf n})  \nonumber \\
  & -\left( c^{\dagger}_s({\bf n + a}_1^{\gamma})(\sigma_x)_{ss'} c_{s'}({\bf n})  +  h.c. \right)   \\
  & - \left( c^{\dagger}_s({\bf n + a}_2^{\gamma})(\sigma_x -i\sigma_z)_{ss'} c_{s'}({\bf n})  + h.c. \right)  \nonumber \\
  & - \left( c^{\dagger}_s({\bf n + a}_3^{\gamma})(\sigma_x + it\sigma_y)_{ss'} c_{s'}({\bf n}) + h.c. \right).  \nonumber
\end{align}
Here ${\bf n} = n_1{\bf a}_1^{\gamma} + n_2{\bf a}_2^{\gamma} + n_3 {\bf a}_3^{\gamma}$  with $n_i$ (i=1, 2, 3) are integers,  and  ${\bf a}_i^{\gamma}$ are the primitive lattice vectors. For the top slab $\gamma\equiv t$,   $n_3$ runs over  the range $(0, \infty)$ and  for the bottom slab $\gamma\equiv b$,   $n_3$ runs over the range $(0, -\infty)$.  %and term $(\sigma_x + it\sigma_y)$ is to be replaced with $(\sigma_x - it\sigma_y)$ . 

The periodicity at the interface is that of the SL, so we need to express the operators in terms of the SL  site labels. So we rewrite the  operators as follows:  $ c_s({\bf n}) = c_s({\bf R} + {\boldsymbol \delta}_{\alpha}, n_3) \equiv d_{\alpha s}({\bf m}, n_3)$ for the top slab ($n_3\geq 0 $)  and  $ c_s({\bf n}) = c_s({\bf R} + {\boldsymbol \delta}_{\alpha}, n_3) \equiv f_{\alpha s}({\bf m}, n_3)$  for the bottom slab ($n_3\leq 0$). Here $\alpha $ is the sublattice index and ${\bf R}= m_1{\bf b}_1 + m_2{\bf b}_2 $ is the position vector of   SL sites (see Fig. \ref{fig-SL}).  By inspection, it is clear that there are geometric symmetries of the model that
involve rotation of the lattice by $\pi$ around both diagonals ( $i.e.$,
the sites in black and red  in Fig. \ref{fig-SL} get interchanged). We will now see how these symmetries are implemented in terms of 
$d_{\alpha s}({\bf m}, n_3)$  and  $f_{\alpha s}({\bf m}, n_3)$.  

First, let us consider the symmetry transformation associated with the $\pi$ rotation about $m_1=m_2$ diagonal.
\begin{align}
\textrm{(i)} \hspace{0.2cm} & U_{1} d_{\alpha s} (m_1, m_2, n_3) U^{\dagger}_{1} = (\sigma_x)_{ss'} f_{\alpha s'} (m_2, m_1, n_3) \nonumber \\
& U_1 i U^{\dagger}_1 = i, \hspace{0.3cm}  U_1 V U_1^{\dagger} = \sigma_x V^{\dagger} \sigma_x 
\end{align}
where the symmetry is unitarily realised and of the particle-particle type, and 
\begin{align}
\textrm{(ii)} \hspace{0.2cm} & Q_1 d_{\alpha s} (m_1, m_2, n_3) Q^{\dagger}_1 = (i\sigma_y)_{ss'} f^{\dagger}_{\alpha s'} (m_2, m_1, n_3) \nonumber \\
& Q_1 i Q_1^{\dagger} = i, \hspace{0.3cm}  Q_1 V Q_1^{\dagger} = -\sigma_y V^{T} \sigma_y
\end{align}
where again, the symmetry is unitarily realised, but is of the Boguliobov or particle-hole type.

Note that the spatial arguments  of  the fermion operators are not the same on both the sides of the equation - in fact
because the  spatial transformation involves a  rotation of  the lattice by $\pi$ about  the diagonal,  $m_1\leftrightarrow m_2$. 
After lengthy algebraic manipulations,  it can be shown  that both the symmetry 
transformations interchange  $H^t$  and  $H^b$  - $i.e.$,  $U_1 H^t U_1^{\dagger} = H^b$  and $Q_1 H^t Q_1^{\dagger}= H^b$, and  vice-versa. 
So if we require both the transformations to be symmetries of the Hamiltonian, then we need the tunnelling matrix to  be of the form $V = V_x \sigma_x + iV_y \sigma_y$  where $V_i (i=x, y)$ are real numbers (assuming also that $V$ has to be  real). This is clear since 
the transformation (i) will be a symmetry of the total Hamiltonian $H = H^t + H^b + H_V$, only if the tunnelling matrix $V$ satisfies  the condition
\begin{align}\label{eq:+ve1}
 U_1 V U_1^{\dagger} = \sigma_x V^{\dagger} \sigma_x  = V, 
\end{align}
 and  the transformation (ii) will  be a symmetry of H only if $V$ obeys the condition
  \begin{align}\label{eq:+ve2}
  Q_1 V Q_1^{\dagger} = -\sigma_y V^{T} \sigma_y = V.
  \end{align}

There are also symmetry transformations associated with $\pi$ rotation around the other diagonal $m_2=-m_1$ of the SL. 
In this case we have the following anti-unitary symmetry transformations:
 \begin{align}
\textrm{(iii)} \hspace{0.2cm} & U_2 d_{\alpha s} (m_1, m_2, n_3) U_2^{\dagger} \nonumber \\
   & = (S)_{\alpha \alpha'} (\sigma_x)_{ss'} f_{\alpha' s'} (-m_2,- m_1, n_3) \nonumber \\
& U_2 i U_2^{\dagger} = -i, \hspace{0.3cm}  U_2 V U_2^{\dagger} = \sigma_x V^{T} \sigma_x 
\end{align}
of the particle-particle type and
\begin{align}
\textrm{(iv)} \hspace{0.2cm} & Q_2 d_{\alpha s} (m_1, m_2, n_3) Q_2^{\dagger} \nonumber \\
   & = (S)_{\alpha \alpha'} (i\sigma_y)_{ss'} f^{\dagger}_{\alpha' s'} (-m_2,- m_1, n_3) \nonumber \\
& Q_2 i Q_2^{\dagger} = -i, \hspace{0.3cm}  Q_2 V Q_2^{\dagger} = -\sigma_y V^{\dagger} \sigma_y. 
\end{align}
of the Boguliobov type. Note also that here the geometric symmetry interchanges $m_1 \leftrightarrow -m_2$.
The matrix  in the sublattice index $S_{\alpha \alpha'}$ has the following non zero elements  $S_{00}=1, S_{13}=S_{31}=1, S_{24}=S_{42}=1$. Both the above transformations  takes $H^t$ to $H^b$ and vice versa. So the transformation (iii) will be a symmetry of $H = H^t + H^b + H_V$, if the tunnelling matrix satisfies the condition 
\begin{align}\label{eq-symmetryiii}
 U_2 V U_2^{\dagger} = \sigma_x V^{T} \sigma_x  = V, 
\end{align}
and  the transformation (iv) will  be a symmetry if
  \begin{align}\label{eq-symmetryiv}
  Q_2 V Q_2^{\dagger} = -\sigma_y V^{\dagger} \sigma_y = V.
  \end{align}

For a real tunnelling matrix, it is clear from  Eq. \ref{eq-symmetryiii} and   Eq. \ref{eq-symmetryiv} that  $V$ needs to  be of the following form  $V = V_x \sigma_x + i V_y \sigma_y$, where $V_x$  and  $V_y$ are real for both symmetries to be realised. Note that this form of V is identical to what was needed for the Hamiltonian to be symmetric under the transformations (i) and (ii). So the symmetry of  rotation by $\pi$ around  either diagonal leads to the same conditions on the tunnelling matrix.

{For commensurate twist angles of the form $\theta_{n_0} =
  \tan^{-1}(1/n_0)$, the geometric symmetry of the super-lattice (SL)
  by the rotation $\pi$ around both the x and y axes will be a
  symmetry of the Hamiltonian $(H^t + H^b)$. For the particular twist
  $n_0=2$, the resulting SL has been shown shown in
      Fig. \ref{fig-SL}. }

{ Consider a rotation by $\pi$ around the x-axis (which passes through
  the centre of the SL unit cell), which takes an SL site $(m_1, m_2)$
  to $(m_1, -m_2+1)$. Clearly the lattice sites in red get
  interchanged with the lattice sites in black.  The sites in red
  (inside the SL unit cell) labelled 0, 1, 2, 3 and 4 get mapped to
  the sites in black labelled 0, 4, 1, 2, and 3 respectively (see
  Fig. \ref{fig-SL}). {This geometric symmetry is realised via the
    following symmetry transformations; }
\begin{align}
\textrm{(i)} \hspace{0.2cm}  & R_{x} d_{\alpha s} (m_1, m_2, n_3) R^{\dagger}_{x} \\ \nonumber
 & = W^{x}_{\alpha \alpha'} (\sigma_x)_{ss'} f_{\alpha s'} (m_1, -m_2 + 1, n_3) \nonumber \\
& R_x i R^{\dagger}_x = i, \hspace{0.3cm}  R_x V R_x^{\dagger} = \sigma_x V^{\dagger} \sigma_x 
\end{align}
where the symmetry is unitarily realised and of the particle-particle type, and 
\begin{align}
\textrm{(ii)} \hspace{0.2cm} & S_x d_{\alpha s} (m_1, m_2, n_3) S^{\dagger}_x \\ \nonumber
& = W^{x}_{\alpha \alpha'} (i\sigma_y)_{ss'} f^{\dagger}_{\alpha s'} (m_1, -m_2 + 1, n_3) \nonumber \\
& S_x i S_x^{\dagger} = i, \hspace{0.3cm}  S_x V S_x^{\dagger} = -\sigma_y V^{T} \sigma_y
\end{align}
where again, the symmetry is unitarily realised, but is of the
Boguliobov or particle-hole type. Since the sites in red labelled 0,
1, 2, 3 and 4 get mapped to the sites in black labelled 0, 4, 1, 2,
and 3 respectively, the symmetric and unitary matrix $W^x_{\alpha
  \alpha'}$ has the following non zero elements : $W^x_{00}=1,
W^x_{14}=W^x_{41}=1, W^x_{21}=W^x_{12}=1, W^x_{32}=W^x_{23}=1,
W^x_{43}=W^x_{34}=1$. Here both the transformations take $H^t \to H^b$
and vice versa.  Imposing the above symmetries on
    the tunnelling Hamiltonian $H_V$, leads to conditions which are
    identical to the conditions
    Eqs. \ref{eq:+ve1}-\ref{eq:+ve2}. Consequently, for a real
  tunnelling matrix, V has to be of the same form that we had earlier
  obtained for symmetry under rotation about the diagonals, where the
  twist angle was $\theta_{n_0} = \pi/2 + tan^{-1}(1/n_0)$. }\\

{Now consider rotation by $\pi$ about the y-axis (which again passes through the centre of the SL unit cell), which  takes an arbitrary SL site $(m_1, m_2)$ to $(-m_1+1, m_2 )$. Clearly the sites in red (inside the SL unit cell) labelled 0, 1, 2, 3 and 4 get mapped to the sites in black labelled 0, 2, 3, 4, and 1 respectively (see Fig. \ref{fig-SL}). {{Again this geometric symmetry is  realised through the following anti-unitary symmetry transformations - }}
\begin{align}
\textrm{(iii)} \hspace{0.2cm}  & R_{y} d_{\alpha s} (m_1, m_2, n_3) R^{\dagger}_{y} \\ \nonumber
 & = W^{y}_{\alpha \alpha'} (\sigma_x)_{ss'} f_{\alpha s'} (-m_1+1, m_2, n_3) \nonumber \\
& R_y i R^{\dagger}_y = - i, \hspace{0.3cm}  R_y V R_y^{\dagger} = \sigma_x V^{T} \sigma_x 
\end{align}
where the symmetry is anti-unitarily realised and of the particle-particle type, and 
\begin{align}
\textrm{(iv)} \hspace{0.2cm} & S_y d_{\alpha s} (m_1, m_2, n_3) S^{\dagger}_y \\ \nonumber
& = W^{y}_{\alpha \alpha'} (i\sigma_y)_{ss'} f^{\dagger}_{\alpha s'} (-m_1+1, m_2, n_3) \nonumber \\
& S_y i S_y^{\dagger} =  - i, \hspace{0.3cm}  S_y V S_y^{\dagger} = -\sigma_y V^{\dagger} \sigma_y
\end{align}
where again, the symmetry is anti-unitarily realised, but is of the Boguliobov or particle-hole type. The symmetric and  unitary matrix $W^y_{\alpha \alpha'}$ has the following non zero elements  $W^y_{00}=1,  W^y_{12}=W^y_{21}=1,   W^y_{23}=W^y_{32}=1,  W^y_{34}=W^y_{43}=1,   W^y_{41}=W^y_{14}=1$. As before, both the transformations (iii) and (iv)  takes  $H^t \to H^b$  i.e. $R_y H^t R_y^{-1} = H^b$  and   $S_y H^t S_y^{-1} = H^b$ and   vice versa.  Now if we impose the above symmetry transformations (iii) and (iv) on  the tunnelling Hamiltonian,  we get conditions on V, which are  exactly identical to the conditions given in Eq. \ref{eq-symmetryiii}-\ref{eq-symmetryiv}. \\
To summarise,  for twist angles of the form  $\theta_{n_0} = tan^{-1}(1/n_0)$,  the geometric symmetries (which lead to symmetries  of H)  are   the rotations by $\pi$ about the  x and y axes, whereas  for   twist angles of the form  $\theta_{n_0} = \pi/2 + tan^{-1}(1/n_0)$,  the geometric symmetries (which lead to symmetries of H)  are  the rotations by $\pi$ about the diagonals  of SL unit cell.  But both the  cases lead to  the same conditions on the tunnelling matrix , $i.e.$,  $V$ has be of the form  $V = V_x \sigma_x + iV_y \sigma_y$ (for a real V), where  $V_x$ and  $V_y $ are  real.}

%----------------------------------------------------------------------------------------------------------------------------------------------------------------------------

%\textcolor{magenta}{\section{Stability under symmetry breaking  perturbation of tunnelling matrix} }
\section{Stability of FA reconstruction under perturbations breaking the symmetries of the $\kappa=0$ Hamiltonian}
\label{App:SymmetryPerturbation}

\begin{figure}[t]
\includegraphics[width=1\linewidth, height=0.45\linewidth]{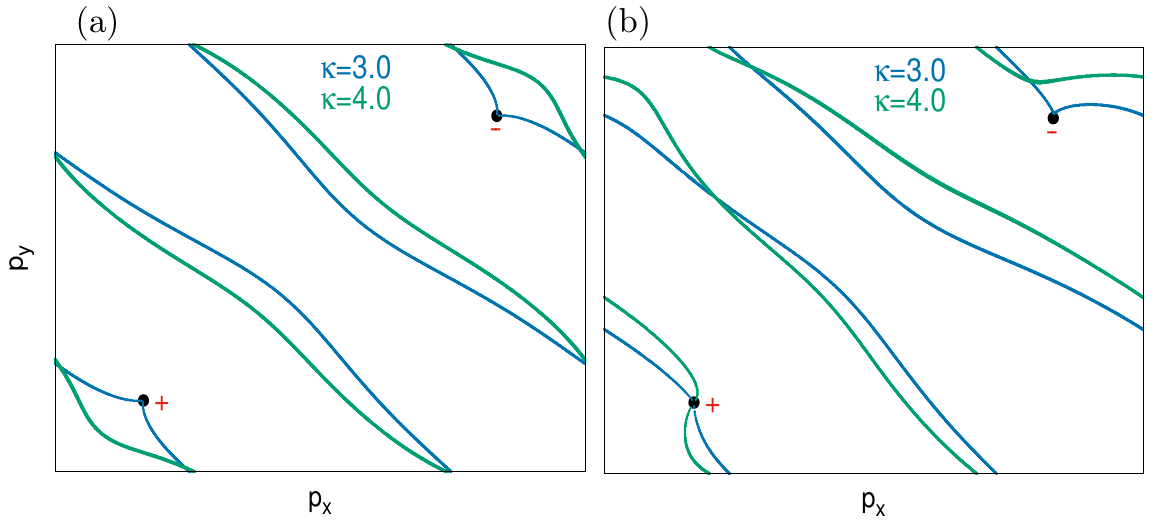}\\
\includegraphics[width=1\linewidth, height=0.45\linewidth]{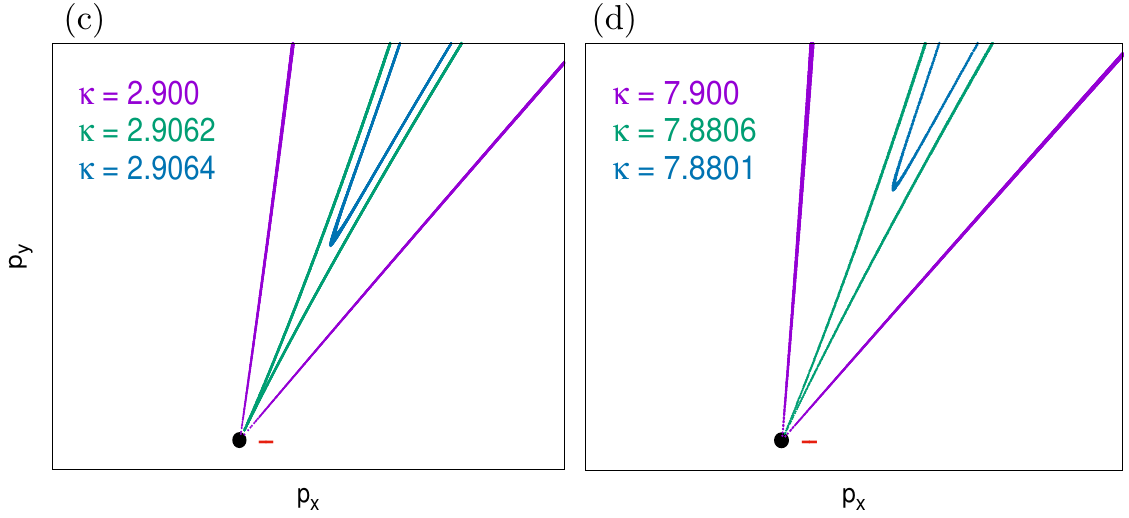}
\caption{$n_0=2$: {The reconstructed Fermi arcs (for $\kappa$ = 3 and 4)  near the lift-off transitions without perturbations in (a)  are compared with the reconstructed Fermi arcs after adding the symmetry breaking  perturbations  in (b). The parameters are given by  $\delta V = V_0 \sigma_0 + V_z\sigma_z$, where $V_0=0.2$ and $V_z = 0.3$. The other parameters are the same as in  Figs. \ref{fig-n=2}(d-f) and in Fig. \ref{fig-n2shape}. Note that symmetry of the Fermi arcs about both the diagonals $p_x=\pm p_y$ is broken in the perturbed case.  Note also that  the perturbed Fermi arcs get detached only from the -ve chiral WPP,  but not from the +ve chiral WPP. (c) and (d) show the singular shapes of the Fermi arcs at and near the  lift-off and  re-attachment transitions, respectively, in the presence of the perturbation. It can be seen that the  singular structure of the Fermi arc  at and near the lift-off and   re-attachment transitions  remains intact.( See also Fig .\ref{fig-n2shape} for the unperturbed case.)  }}
\label{fig-n2op}
\end{figure}

{ Here, we show that adding perturbations that break the symmetries of
  the $\kappa=0$ Hamiltonian do not change the results qualitatively.
  To see this, we consider a particular twist $n_0=2$ with overlapping
  Weyl point projections. Recall that for $V=\sum_{\mu=0}^{4}
  V_{\mu}\sigma^{\mu}$ the symmetries forced $V_0,V_z=0$ and $V_x,V_y$
  real.  We will break these symmetries by allowing for nonzero
  $V_0,V_z$. For small couplings $\kappa<1$, the reconstructed Fermi
  arcs look very similar to those shown in Fig. \ref{fig-n=2}(a). To
  check whether the lift-off and re-attachment transitions retain the
  singular shape of the Fermi arc close to the transitions, we study
  the reconstructed Fermi arcs, close to the transitions, with the
  symmetry breaking perturbations. The comparison with the earlier
  unperturbed results is shown in Fig. \ref{fig-n2op} and we find that
  the singular shape of the Fermi arc at and near the lift-off and
  re-attachment survices adding symmetry breaking perturbations to the
  tunnelling matrix.  With the particular choice of $V_{\mu}$ we have
  made, the Fermi arc gets detached only from the negative chirality
  WPP, but remains attached to the positive chirality WPP.  However,
  we have confirmed that various choices of $V_\mu$ can lead to
  detachment from either or both of the WPPs. }

%----------------------------------------------------------------------------------------------------------------------------------------------------------------------------

\section{Computation of Fermi Arc States at the Junction}
\label{App:ComputationFermiArc}
In this section we will provide  the details for the computation of the interface localized states. Our computation will closely follow Ref. \onlinecite{Murthy2020}.

\subsection{Surface states of the slabs}
The idea is to look for decaying eigenstates into the bulk for  both the slabs. Then we match solutions at the interface via the  coupling term  $H_V$ to get the interface localised states. Translational invariance is broken in the $z$-direction,  but it remains unbroken in the transverse directions. The eigenstates can, 
hence, be labeled by the momenta $(k_x, k_y)$.  We proceed with the following  ansatz for the decaying eigenstates,  -

\begin{align}
\textrm{top slab:} \hspace{0.2cm} \ket{E, k_x, k_y}^t &= \sum_{n=0}^{\infty} \sum_s u_t^{n} {\bf \Phi}_{s}^t c^{\dagger}_{ns}\ket{0}\\
\textrm{bottom slab:} \hspace{0.2cm} \ket{E, k_x, k_y}^b &= \sum_{n=0}^{-\infty} \sum_s u_b^{-n} {\bf \Phi}_{s}^b c^{\dagger}_{ns}\ket{0}
\end{align}
where $|u_{\gamma}|<1$, for the state $\ket{E, k_x, k_y}$ to be normalizable and $n$ is the discretised $z$ coordinate and here,  ${\gamma}=(t, b)$. We then solve the Schrodinger equation, $H^{\gamma} \ket{E, k_x, k_y}^{\gamma}=E \ket{E, k_x, k_y}^{\gamma}$ to obtain  $(u_{\gamma}, \Phi^{\gamma}_s)$. As shown in  Ref. \onlinecite{Murthy2020},  there are two normalizable solutions for $u_{\gamma}$ for a given $E$ and ${\bf k}=(k_x, k_y)$,  which  are given by

\begin{align} \label{eq-u}
(u_{\gamma})_{\pm} &=  \frac{1}{2}(\xi_{\gamma} \pm \sqrt{\xi_{\gamma}^2 - 4}), 
\end{align}
where,  $\xi_{\gamma}$ is a solution of the following  quadratic equation,
\begin{align} \label{eq-xi}
(1 - t'^2) \xi_{\gamma}^2 - 4 f_1^{\gamma} \xi_{\gamma}  -  E^2 + 4( (f_1^{\gamma})^2 + (f_3^{\gamma})^2 + t'^2)=0.
\end{align}
%\begin{align}
%(\xi_{\alpha})_{\pm} = \frac{2f_1^{\alpha} \pm \sqrt{(2f_1^{\alpha})^2 + (1 - t'^2)(E^2 - 4( (f_1^{\alpha})^2 + (f_3^{\alpha})^2 + t'^2) )  }   }{1 - t'^2  }
%\end{align}
For each root of $\xi_{\gamma}$, it is obvious from Eq. \ref{eq-u} that  the two roots of $u_{\gamma}$ obey  $(u_{\gamma})_{+}(u_{\gamma})_{-}=1$. This implies that one of the roots of $u_{\gamma}$ for each root of $\xi_{\gamma}$ must satisfy $|u_{\gamma}| \leqslant 1$. The equality holds only when $\xi_{\gamma}$ is real and $-2 \leqslant \xi_{\gamma} \leqslant 2$. Therefore two roots of $\xi_{\gamma}$ give two solutions of $u_{\gamma}$ which give  normalizable decaying eigenstates solutions for both slabs, only when  $\xi_{\gamma}$ (which may be complex) lies outside the range   $-2 \leqslant \xi_{\gamma} \leqslant 2$.  

A few further steps of algebra suffices to show that the spin wavefunctions ${\bf \Phi}^{\gamma}$ can be expressed as 
\begin{align}
{\bf \Phi}^t &= \left(1,  \frac{-(2f^t_3 - E)}{2f_1^t - (u_t+1/u_t) + t'(u_t-1/u_t)  } \right)^T \\
{\bf \Phi}^b &= \left(\frac{(2f^b_3 + E)}{2f_1^b - (u_b+1/u_b) + t'(u_b-1/u_b)}, 1 \right)^T.
\end{align}

We can now  construct a general wavefunction for the  top slab as 
\begin{align}
\ket{\Psi^t({\bf k})} =& \sum_{n=0}^{\infty}  A_1^t({\bf k}) (u_1^t({\bf k}))^n \Phi^t_1({\bf k})c^{\dagger}_n({\bf k})\ket{0} \nonumber \\
& \hspace{0.6cm}  +  A_2^t({\bf k}) (u_2^t({\bf k}))^n \Phi^t_2({\bf k}) c^{\dagger}_n({\bf k})\ket{0} \nonumber \\
=& \sum_{n=0}^{\infty} \psi^t_n({\bf k}) c^{\dagger}_n({\bf k})\ket{0} ,
\end{align}
and for the bottom slab as 
\begin{align}
\ket{\Psi^b({\bf k})} =& \sum_{n=0}^{-\infty}  A_1^b({\bf k}) (u_1^b({\bf k}))^{-n} \Phi^b_1({\bf k})  c^{\dagger}_n({\bf k})\ket{0}  \nonumber \\
 & \hspace{0.6cm} +  A_2^b({\bf k}) (u_2^b({\bf k}))^{-n} \Phi^b_2({\bf k})  c^{\dagger}_n({\bf k})\ket{0} \nonumber \\
=& \sum_{n=0}^{-\infty} \psi^b_n({\bf k}) c^{\dagger}_n({\bf k})\ket{0},
\end{align}
where,
\begin{align}
  \psi^t_n({\bf k})  = &  A_1^t({\bf k}) (u_1^t({\bf k}))^n \Phi^t_1({\bf k})   +  A_2^t({\bf k}) (u_2^t({\bf k}))^n \Phi^t_2({\bf k}), \nonumber \\
  \psi^b_n({\bf k}) =  & A_1^b({\bf k}) (u_1^b({\bf k}))^{-n} \Phi^b_1({\bf k})  + A_2^b({\bf k}) (u_2^b({\bf k}))^{-n} \Phi^b_2({\bf k}) , 
 \label{eq-layer-wavefn2}  
 \end{align} 
  and    $A_1^{\gamma}$, $A_2^{\gamma}$ are unknown  constants. Next, our goal is to match the wavefunctions at the interface through the coupling term $H_V$ to fix the constants $A_1^{\gamma}$, $A_2^{\gamma}$.

%-----------------------------------------------------------------------------------------------------------------------------------------------------------------

\subsection{Manipulation of coupling term $H_V$}
Before proceeding further, we first need to  write the coupling term $H_V$ in the transverse momentum space ${\bf k}=(k_x, k_y)$. On the interface, the true periodicity is that of superlattice. We label the operators which live on the zeroth layer of the top and bottom slabs by 
$ \hspace{0.2cm} d_{\alpha s}({\bf R}), \hspace{0.2cm} \textrm{and} \hspace{0.2cm} f_{\alpha s}({\bf R}) $ respectively, where  the superlattice sites are given by ${\bf R}= m_1{\bf b_1} + m_2 {\bf b_2}$, $m_i \in \mathbb{Z}$. Given a superlattice (SL) site, we then assign a  `star' of sites in both slabs as sublattice sites to
this SL site with  `$\alpha$' representing the sublattice index. Writing the term $H_V$ in the SL index, we  then get the following  form from Eq. \ref{eq-coupling} in the. main paper,
\begin{align}\label{eq-hvsl1}
H_V= &  \sum_{{\bf R}, \alpha} \sum_{{\bf R'}, \beta} \{  d^{\dagger}_{\alpha s}({\bf R}) V_{ss'}(|{\bf R - R'} + {\boldsymbol \delta}_{\alpha}^t - {\boldsymbol \delta}_{\beta}^b |) f_{\beta s'} ({\bf R'}) \nonumber \\
 & \hspace{2cm} + H.c \}, 
\end{align}
where we have used  $ c_{t0s}({\bf r}_t) = c_{t0s}({\bf R}+{\boldsymbol \delta}^t_{\alpha}) \equiv  d_{\alpha s}({\bf R}) $ and $ c_{b0s}({\bf r}_b) = c_{b0s}({\bf R}+ {\boldsymbol \delta}^b_{\beta})  \equiv f_{\beta s}({\bf R})$. We Fourier transform the operators,
\begin{align}\label{eq-ft_d}
d_{\alpha s}({\bf R}) = & \frac{1}{\sqrt{N_{SL}}} \sum_{{\bf p}\in BZ_{SL}}  e^{i{\bf p . R}} \tilde{d}_{\alpha s}({\bf p}) \nonumber \\
f_{\beta s}({\bf R}) = & \frac{1}{\sqrt{N_{SL}}} \sum_{{\bf p}\in BZ_{SL}}  e^{i{\bf p . R}} \tilde{f}_{\beta s}({\bf p}),
\end{align}
where $N_{SL}$ is total number of SL lattice sites and substitute these expressions  in Eq. \ref{eq-hvsl1},  to  get the following  coupling term after simplification,
\begin{align}\label{eq-hvsl2}
H_V= &  \sum_{{\bf p}\in BZ_{SL}} \sum_{\alpha \beta} \{ \tilde{d}_{\alpha s}^{\dagger}({\bf p}) \tilde{V}_{ss'}^{\alpha \beta}({\bf p})  \tilde{f}_{\beta s'}({\bf p}) + H.c \} , \\
%\end{align}
 \textrm{ where} &  \nonumber \\
% \begin{align}
  &  \tilde{V}_{ss'}^{\alpha \beta}({\bf p}) =  \sum_{\bf R} e^{-i{\bf p . R}} V_{ss'}({\bf R} + {\boldsymbol \delta}_{\alpha}^t - {\boldsymbol \delta}_{\beta}^b ).
  \label{eq-v(p)}
 \end{align}
 
The ${\boldsymbol \delta}_{\alpha}^t$ and ${\boldsymbol \delta}_{\beta}^b$ are the relative position vectors of the sublattice sites  with respect to the SL unit cell. Since the states $\ket{\Psi^t({\bf k}) }$ (${\bf k} \in BZ_t$) and $\ket{\Psi^b({\bf k}) }$ (${\bf k} \in BZ_b$) are defined where the operators $c_{ns}^{\dagger}({\bf k}) $ act on the vacuum $\ket{0}$, we need to rewrite Eq. \ref{eq-hvsl2} in terms of the operators $c_{t0s}^{\dagger}({\bf k})$ and $c_{b0s}^{\dagger}({\bf k})$. 
So we want to relate the Fourier transforms of $d_{\alpha s}({\bf R})$  and $f_{\alpha s}({\bf R})$ with that of $c_{t0s}({\bf r}_t)$  and $c_{b0s}({\bf r}_b)$. 
We recall that
\begin{align}\label{eq-d_c}
d_{\alpha s}({\bf R}) &= c_{t0s}({\bf R} + \boldsymbol{\delta}_{\alpha}^t) \nonumber \\
 & =  \frac{1}{\sqrt{N}} \sum_{{\bf k} \in BZ_t} e^{i {\bf k.(R} + \boldsymbol{\delta}_{\alpha}^t)} c_{t0s}({\bf k}) \nonumber \\
 & =  \frac{1}{\sqrt{N}} \sum_{{\bf p} \in BZ_{SL}} \sum_{l=1}^{N_{sc}/2}  e^{i ({\bf p + Q}_l).({\bf R} + \boldsymbol{\delta}_{\alpha}^t)} c_{t0s}({\bf p + Q}_l) \nonumber \\
  = &  \frac{1}{\sqrt{N_{SL}}} \sum_{{\bf p} \in BZ_{SL}}  \frac{e^{i{\bf p.R}}}{\sqrt{N_{sc}/2} }   \sum_{l=1}^{N_{sc}/2}  e^{i( {\bf p+ Q}_l).\boldsymbol{\delta}_{\alpha}^t} c_{t0s}({\bf p + Q}_l).
\end{align}
In going from the second to the third step, we have used the fact that the momenta ${\bf k} \in BZ_t$ can be decomposed as:
${\bf k}_l = {\bf p + Q}_l$, where ${\bf p} \in BZ_{SL}$  and ${\bf Q}_l = n_1 {\bf g}_1 + n_2 {\bf g}_2$, $n_i \in \mathbb{Z}$.
 Since ${\bf g}_1$ and ${\bf g}_2$ are the reciprocal lattice vectors of the SL lattice, the following holds  ${\bf Q}_l.{\bf R}=2\pi \times (integer)$. 
 Here the label $l$  goes over the first and second Brillouin zones, since  the SL BZ  is $N_{sc}/2$ times smaller than the 
 original BZ.
 From  the third to the fourth step,  we use the identity $e^{i{\bf Q}_l.{\bf R}} =1$ and write $N=N_{SL}\times N_{sc}/2$. Recall  that the number of sites per SL unit cell is given by  $N_{sc}=2\times (n_0^2 + 1)$ when $n_0$ is even and by $N_{sc}= (n_0^2 + 1)$ when $n_0$ is odd. In particular when $n_0=2$ (see Fig. \ref{fig-SL-BZ} in the main paper), there are 5 sites of the top layer and 5 sites of the bottom layer per SL unit cell and a set of 5  values of   ${\bf Q}_l =({\bf 0,  g_1,  g_2,  -g_1, -g_2})$. Comparing  Eq. \ref{eq-d_c} with the Eq. \ref{eq-ft_d}, we get the following   relations - 
\begin{align}
\tilde{d}_{\alpha s}({\bf p}) =   \frac{1}{\sqrt{N_{sc}/2} } \sum_{l=1}^{N_{sc}/2}  e^{i( {\bf p+ Q}_l).\boldsymbol{\delta}_{\alpha}^t} c_{t0s}({\bf p + Q}_l).
\end{align} 
and similarly for $\tilde{f}_{\beta s}({\bf p}) $, we get 
\begin{align}
\tilde{f}_{\beta s}({\bf p}) =   \frac{1}{\sqrt{N_{sc}/2} } \sum_{l=1}^{N_{sc}/2}  e^{i( {\bf p+ Q}_l).\boldsymbol{\delta}_{\beta}^b} c_{b0s}({\bf p + Q}_l).
\end{align} 
Now substituting  these expression of $\tilde{d}_{\alpha s}({\bf p}) $ and $\tilde{f}_{\beta s}({\bf p})$ in Eq. \ref{eq-hvsl2}, we  finally  get the following  
expression for the coupling term
\begin{align}\label{eq-Hv_final}
H_V = & \frac{1}{N_{sc}/2} \sum_{{\bf p} } \sum_{ss'} \sum_{\alpha, \beta =0}^{N_{sc}/2-1} \hspace{0.2cm} \sum_{l, l'=1}^{N_{sc}/2}  \{ e^{-i {\bf Q}_l.\boldsymbol{\delta}_{\alpha}^t + i {\bf Q}_{l'}.\boldsymbol{\delta}_{\beta}^b }  \nonumber  \\
 & \hspace{0.2cm} \times c^{\dagger}_{t0s}({\bf p + Q}_l) \tilde{V}_{ss'}^{\alpha \beta}({\bf p}) c_{b0s'}({\bf p + Q}_{l'}) + H.c\}, 
\end{align} 
where $\tilde{V}_{ss'}^{\alpha \beta}({\bf p})$ is given in Eq. \ref{eq-v(p)}. \\

%-------------------------------------------------------------------------------------------------------------------------------------------------------------

\subsection{Matching conditions}
Since $\ket{\Psi^t({\bf k}_l)}$ and  $\ket{\Psi^b({\bf k}_l)}$ are the eigenstates of $H^t$ and $H^b$ respectively and  ${\bf k}_l = {\bf p + Q}_l$,  we can write the eigenstates   of $H$ as $\ket{\Psi({\bf k}_l)} = \ket{\Psi^t({\bf k}_l)} + \ket{\Psi^b({\bf k}_l)}$. Note that we have yet to determine the constants $A_1^t, A_2^t$  and  $A_1^b, A_2^b$,  defined in Eq. \ref{eq-layer-wavefn2}. Solving for $H \ket{\Psi({\bf k}_l)} = E \ket{\Psi({\bf k}_l)}$, gives us a discrete Schrodinger equation for a generic layer $n$ written as 
\begin{align}\label{eq-recursive}
h^{\gamma}_{nn} \psi_n^{\gamma}({\bf k}_l)  + h^{\gamma}_{n,n-1} \psi_{n-1}^{\gamma}({\bf k}_l)   + h^{\gamma}_{n, n+1} \psi_{n+1}^{\gamma}({\bf k}_l)   = E \psi_n^{\gamma}({\bf k}_l)
\end{align}
where $\gamma =(t, b)$  and the different $h$ matrices are  obtained from the Eqs. (3) and (4) in the main paper -
\begin{align}
h^{\gamma}_{nn}  = {\bf M}^{\gamma}, \hspace{0.3cm}  h^{\gamma}_{n, n+1} = -{\bf T}^{\dagger} ,  \hspace{0.3cm}  h^{\gamma}_{n, n-1} = - {\bf T}.
%h^b_{nn} & = {\bf M}^b, \hspace{0.3cm}  h^b_{n, n+1} = -{\bf T},  \hspace{0.3cm}  h^b_{n, n-1} = - {\bf T}^{\dagger}. 
\end{align}
(i) Next, let us  consider $\gamma=t$ and $n=0$. Now when we solve $H \ket{\Psi({\bf k}_l)} = E \ket{\Psi({\bf k}_l)}$. In this case, the $n=-1$ layer does not exist and is instead replaced by the $n=0$ layer of the bottom slab.   Hence, the coupling matrix 
$H_V$ will also act on the states $\ket{\Psi({\bf k}_l)}$; this results in  the following equation,
\begin{align}\label{eq-recursive-v}
 h^{t}_{00} \psi_0^{t}({\bf k}_l)  +   \sum_{l'} \tilde{h}(l, l') \psi^b_0({\bf k}_{l'})    + h^{t}_{0, 1} \psi_{1}^{t}({\bf k}_l)   = E \psi_0^{t}({\bf k}_l),   
\end{align}
 where the matrix  $\tilde{h}(l, l')$ is given by  
\begin{align}
   \tilde{h}(l, l') = \frac{\kappa}{N_{sc}/2} \sum_{\alpha \beta} e^{(-i {\bf Q}_l.\boldsymbol{\delta}_{\alpha}^t + i {\bf Q}_{l'}.\boldsymbol{\delta}_{\beta}^b) } V^{\alpha \beta}({\bf p}). 
 \end{align}
Eq. \ref{eq-recursive}  remains valid  even for  $n=0$ (and setting  $\gamma=t$), we get 
\begin{align}\label{eq-recursive-0}
h^{t}_{00} \psi_0^{t}({\bf k}_l)  + h^{t}_{0, -1} \psi_{-1}^{t}({\bf k}_l)   + h^{t}_{0, 1} \psi_{1}^{t}({\bf k}_l)   = E \psi_0^{t}({\bf k}_l),
\end{align}
where  $h^t_{0, -1} = - {\bf T}$. Now we can combine both  Eq. \ref{eq-recursive-v} and Eq. \ref{eq-recursive-0} to  finally get 
the wavefunction  matching condition at the interface:
\begin{align}\label{eq-match1}
- {\bf T} \psi_{-1}^{t}({\bf k}_l)  +   \sum_{l'} \tilde{h}(l, l') \psi^b_0({\bf k}_{l'})  = 0.
\end{align}
(ii) Similarly we can consider  $\gamma=b$ and $n=0$, and get the second matching condition at the interface:
\begin{align} \label{eq-match2}
- {\bf T}^{\dagger}  \psi_{1}^{b}({\bf k}_l)  +   \sum_{l'} \tilde{h}^{\dagger}(l, l') \psi^t_0({\bf k}_{l'})  = 0.
\end{align}

Notice that  in Eqs.\ref{eq-match1} and \ref{eq-match2}, each term is a  $2\times1$ column  matrix. There are  $N_{sc}/2$ number of 
different ${\bf k}_l= {\bf p + Q}_l$ values. So each equation is essentially  a set of $N_{sc}$ equations. 
Eqs.\ref{eq-match1} and \ref{eq-match2} together give a set of $2 \times N_{sc}$ equations and there are as many unknown constants ($A^{\gamma}_1({\bf k}_l), A^{\gamma}_2({\bf k}_l)$) to be determined (see Eq. \ref{eq-layer-wavefn2}). We can combine  Eq. \ref{eq-match1} and  Eq. \ref{eq-match2} to give a single  matrix equation,
\begin{align}
  M (E, \kappa, p_x,  p_y) A = 0, 
\end{align}
where $A = (A^t _1, A^t_2,  A^b_1, A^b_2)^T$ with,
\begin{align*}
A^t_1 &= (A^t_1(l_1),  A^t_1(l_2), ...A^t_1(l_{N_{sc}/2}))  \\
A^t_2 &= (A^t_2(l_1),  A^t_2(l_2), ...A^t_2(l_{N_{sc}/2}))  \\
A^b_1 &= (A^b_1(l_1),  A^b_1(l_2), ...A^b_1(l_{N_{sc}/2}))  \\
A^b_2 &= (A^t_2(l_1),  A^t_2(l_2), ...A^t_2(l_{N_{sc}/2})).  
\end{align*}
 Non trivial solutions for the constants  lead to exponentially localised states at the interface, and for  non trivial solutions to exist, we must have  $\textrm{det}(M(E, \kappa, p_x, p_y)) = 0$,  for a given $E$ and $\kappa$. The determinant vanishing condition gives an equation, which  $p_x$ and $p_y$ must satisfy for a given $(E, \kappa)$. The set of all such $(p_x, p_y)$, for $E=0$ and fixed $\kappa$,  yields  the reconstructed Fermi arc at the interface. 

%---------------------------------------------------------------------------------------------------------------------------------------------------------------
\section{Twists with $n_0=4$ and $n_0=5$ }
\label{App:n0=4And5}

We have already discussed the results for a non trivial  twist with $n_0=2$ for both the cases of  overlapping and non overlapping WPPs in detail in the main text. 
Here, for completeness,  we consider higher values of $n_0$.  First, we clarify that the case for  $n_0=3$  has not been studied here in detail, because it is very similar to the case with  $n_0=2$  - in both cases, their  superlattices are similar - the SL unit cell contains the same number of lattice sites $N_{sc}=10$  and the unit cell is $N_{sc}/2=5$ times larger than the original unit cell.  So we discuss below, the cases when $n_0=4$ and $n_0=5$.

\begin{figure}[t]
\includegraphics[width=0.9\linewidth, height=0.5\linewidth]{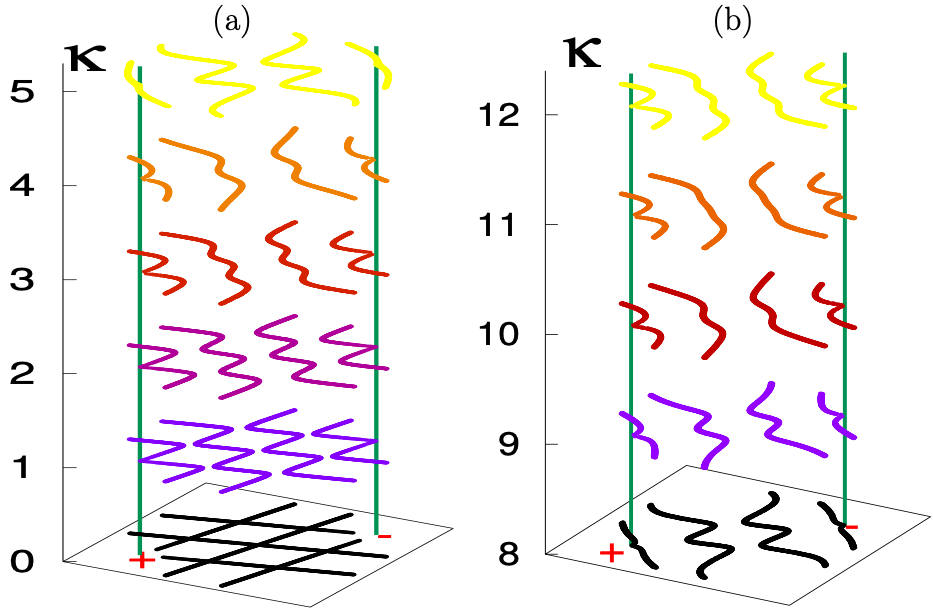}\\
\includegraphics[width=0.9\linewidth, height=0.45\linewidth]{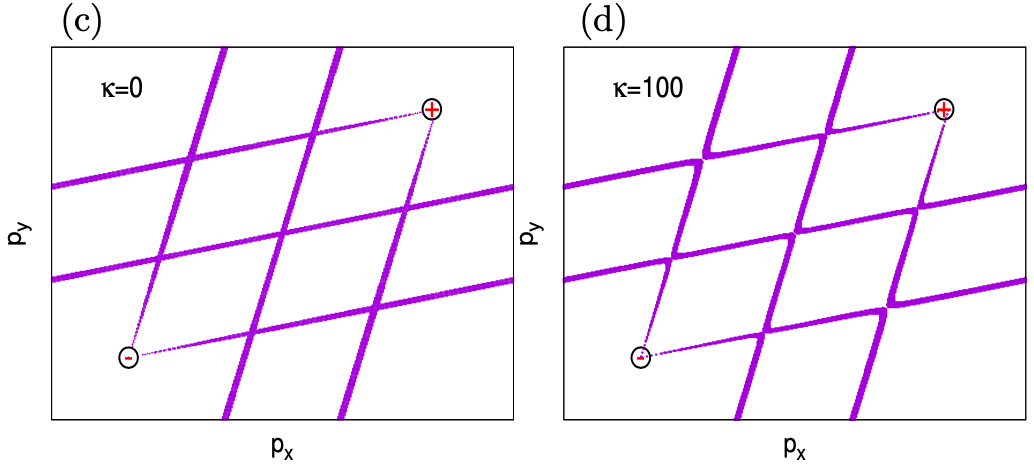}
\caption{$n_0=4$ with overlapping Weyl point projections in the surface BZ. In (a) and (b),  reconstructed Fermi arcs are shown  for couplings  in the ranges $\kappa=(0, 5)$ and $\kappa=(8, 12)$ respectively.  The detachment  of the FAs from the WPPs and its subsequent reattachment occur in the coupling range $\kappa=(5, 7)$. This region of $\kappa$ is explored separately in Fig.6. Parameters taken are  $t'=1.5$  and $V_x =1, V_y=0$. 
The duality between the strong and weak coupling in the  FA reconstruction is shown in (c) and (d). }
\label{fig-n4}
\end{figure}

\textcolor{magenta}{\subsection{Twist with ${\bf n_0=4}$}}
Here, we shall  consider the twist with $n_0=4$ for the case when the   WPPs overlap. 
 We start with a system with the slabs  aligned and the Fermi arcs lying along the $x$-axis. Then the top slab is given a twist by an  angle $\theta_4=\tan^{-1}(1/4)$  counterclockwise and the bottom slab is  twisted  counterclockwise by the angle $(\pi/2 - \theta_4)$,  such that the site $(4, -1)$ of top layer lies on  top of the  site $(1, -4)$ of the bottom layer. We choose $k_0=2\pi/3$ so that the positively charged chiral WPP of the  top slab  lies on the positively charged chiral WPP of the bottom slab and the negatively charged chiral WPPs also lie on top of each other. The overlapping occurs in the first SL BZ after translating the WPPs by the appropriate reciprocal lattice vectors of the SL. 

\begin{figure*}
\includegraphics[width=0.9\linewidth, height=0.45\linewidth]{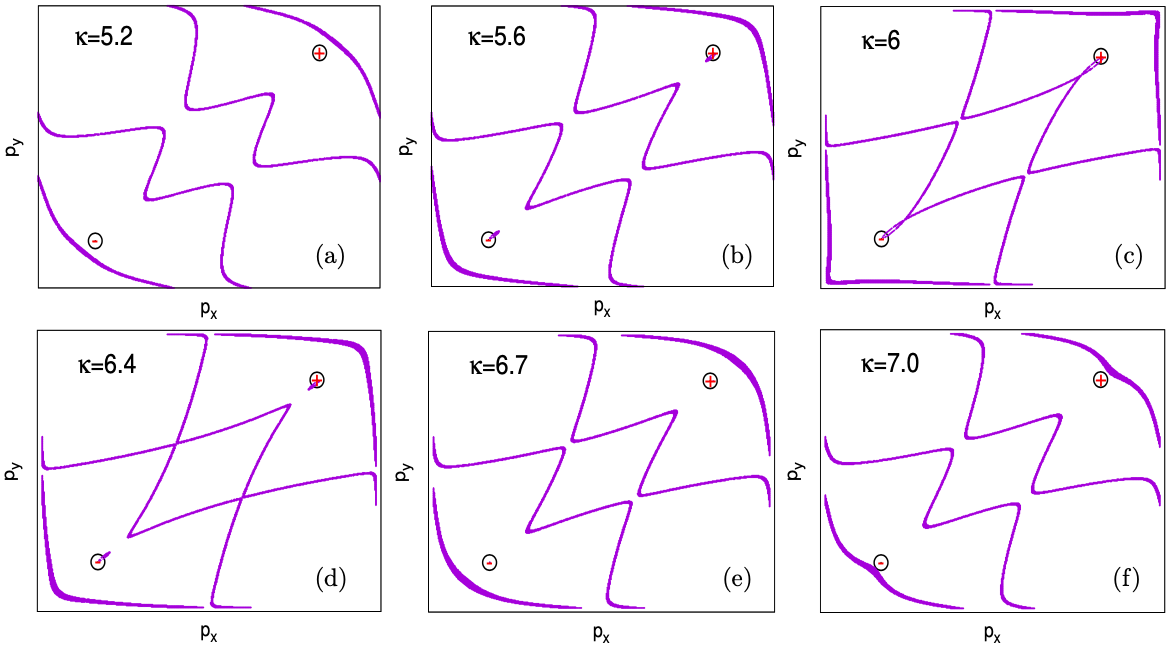}
\caption{The surface BZ for $n_0=4$ in the coupling range   $\kappa=(5.2, 7.0)$  is explored, where the FA  detaches  from the WPPs and  then reattaches  again. }
\label{fig-n4-detach}
\end{figure*}

The reconstructed Fermi arc  is shown in Fig. \ref{fig-n4}.  As in the  $n_0=2$ case, we find a duality in the Fermi arc reconstruction between strong and weak inter-layer coupling strengths. There is also a regime of parameters where  the  Fermi arc  is detached from the WPPs.  The Fermi arc detachment occurs
at  around $\kappa \sim 5$ and then it again re-attaches at $\kappa \sim 7$. This is depicted in Fig. \ref{fig-n4-detach}. 

\begin{figure*}
\includegraphics[width=1\linewidth, height=0.5\linewidth]{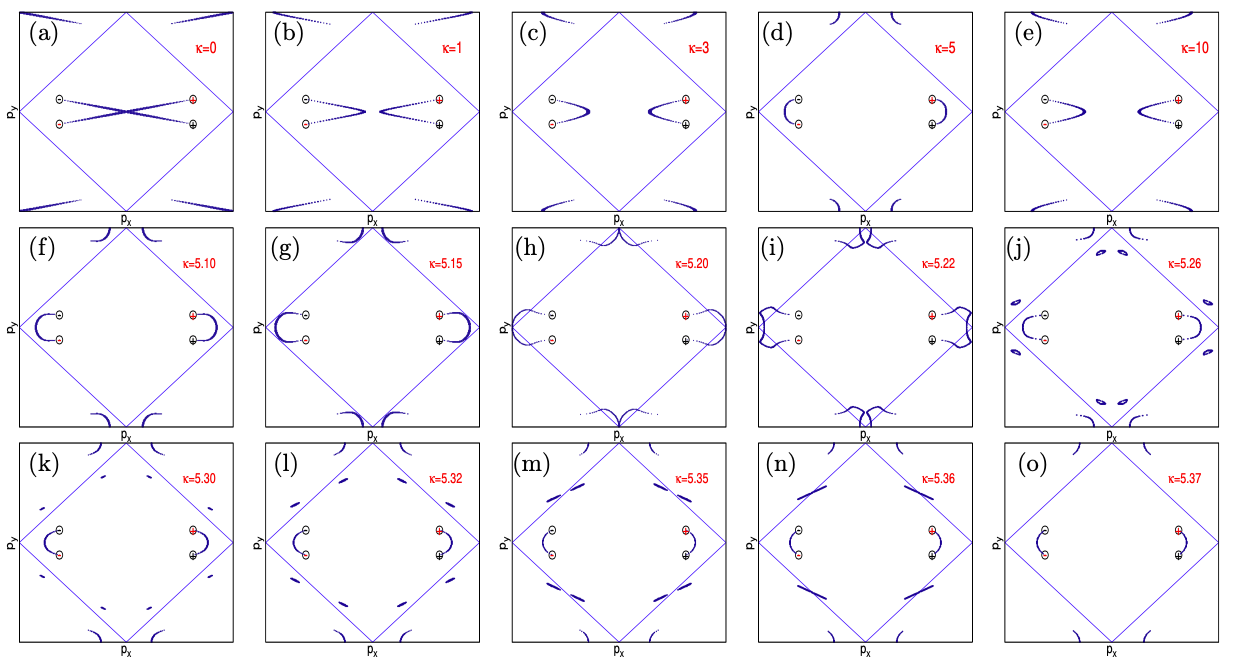}
\caption{$n_0=5$: The panels show the evolution of the  reconstructed Fermi arcs with the  coupling strength $\kappa$. The parameters are  $k_0=\pi/4$,  $t'=0.8$,  $V_x=1, V_y=0$. The inscribed blue square is the boundary of the superlattice BZ. The Weyl point projections of the top (bottom) slab are in red (black) colour. In the first row, the reconstructed Fermi arcs are  shown for $\kappa=0, 1,  3,  5 $ and 10. Notice the sign change of curvature of Fermi arc for $\kappa=5$. In the second and third rows,  the Fermi arc reconstruction around the coupling $\kappa=5$ has been explored in more detail. In the second row,  as $\kappa$ is increased from 5,  the Fermi arc deforms and at  around $\kappa= 5.26$, it splits  to produce a pair of  extra small closed  loops which are  inside the first superlattice  BZ. In the third row  with increasing $\kappa$,  the extra small closed loops move  towards the boundary of the superlattice BZ and merge with the another small loop coming from outside the  first superlattice BZ. Finally they disappear at $\kappa=5.37$, leaving only those Fermi arcs which are attached to the WPPs. At larger couplings,  the reconstructed Fermi arcs  are similar to the small coupling results (e.g. figure (c) and (e) ), as expected from the duality between strong and weak couplings.   }
\label{fig-n=5}
\end{figure*}

\textcolor{magenta}{\subsection{Twist with  ${\bf n_0=5}$}}
The first non trivial case with  $n_0$  odd is when $n_0=5$. Here we consider the Fermi arc reconstruction for the case of non overlapping WPPs. We rotate the top slab counterclockwise  and  the bottom slab clockwise until the lattice site $(5, -1)$ of the  top layer and  the  site $(5, 1)$ of the bottom layer lie on top of each other. Here the angle of rotation is $\theta_5 = \tan^{-1}(1/5)$.  We take $k_0=\pi/4$. The Weyl point projections of positive and negative  chirality are  now at $k_0(\cos{\theta_5}, \pm \sin{\theta_5})$  and  $k_0(-\cos{\theta_5}, \mp \sin{\theta_5})$ respectively,  where the upper (lower) sign is for the  top (bottom) slab. The reconstructed Fermi arcs  are  shown in Fig. \ref{fig-n=5}.  At zero coupling, there are unreconstructed Fermi arcs of  the individual slabs. As we switch on and increase the coupling, the reconstructed Fermi arcs  change their  signs of curvature at around $\kappa=5$ (see Fig. \ref{fig-n=5}(c) and Fig. \ref{fig-n=5}(d) ). The Fermi arcs deform and split when the coupling is increased  slowly beyond $\kappa=5$. After splitting, a pair of small closed loops are  formed. The small closed loops approach  the superlattice BZ boundary and finally disappear after merging. An almost similar feature exists also for the $n_0=2$ case,  where newly formed  extra small loops disappear after merging at the superlattice  BZ boundary (see Fig. \ref{fig-n=2}(c)).

%---------------------------------------------------------------------------------------------------------------------------------------------------------------

%-------------------------------------------------------------------------------------------------------------------------------------------------------------
\bibliographystyle{apsrev}

%----------------------------------------------------------------------------------------------------------------------------------------------------------

\end{document}